\documentclass [11pt]{article}
 \pdfoutput=1
\usepackage{amsfonts}
\usepackage{graphicx}
\usepackage{amsmath,amssymb}
\usepackage{latexsym}
\usepackage{mathrsfs}
\usepackage{bbold}
\usepackage{color}
\usepackage{slashed}
\usepackage{cancel}
\usepackage{soul} %
\usepackage{setspace} %
\usepackage{comment}
\usepackage{booktabs}
\usepackage{multirow}
\usepackage[nosort]{cite}
\usepackage[small]{caption}

\definecolor{red}{rgb}{1,0,0}

\makeatletter
\def\section{\@startsection {section}{1}{\z@}{-3.5ex plus -1ex minus
 -.2ex}{2.3ex plus .2ex}{\large\bf}}
\def\subsection{\@startsection{subsection}{2}{\z@}{-3.25ex plus -1ex
minus -.2ex}{1.5ex plus .2ex}{\normalsize\bf}}
\makeatother
\makeatletter

\@addtoreset{equation}{section}

\makeatother

\def\bea{\begin{eqnarray}} \def\eea{\end{eqnarray}}
\def\be{\begin{equation}} \def\ee{\end{equation}} 
\def\nn{\nonumber}

\usepackage[colorlinks,linkcolor=black,citecolor=blue,urlcolor=blue,linktocpage]{hyperref}

\begin{document}

\thispagestyle{empty}

\begin{center}

	\vspace*{-.6cm}

	\begin{center}

		\vspace*{1.1cm}

		{\centering \Large\textbf{Renormalization Scheme Dependence, RG Flow\\[5pt] and Borel Summability in $\phi^4$ Theories in $d<4$}}

	\end{center}

	\vspace{0.8cm}
	{\bf Giacomo Sberveglieri$^{a}$, Marco Serone$^{a,b}$, and Gabriele Spada$^{c}$}

	\vspace{1.cm}

	${}^a\!\!$
	{\em SISSA and INFN, Via Bonomea 265, I-34136 Trieste, Italy}

	\vspace{.3cm}

	${}^b\!\!$
	{\em ICTP, Strada Costiera 11, I-34151 Trieste, Italy}

	\vspace{.3cm}

	${}^c\!\!$
	{\em Laboratoire Kastler Brossel, ENS - Universit\'e PSL, CNRS, Sorbonne Universit\'e,\\ Coll\`ege de France,  24 Rue Lhomond, 75005 Paris, France}

\end{center}

\vspace{1cm}

\centerline{\bf Abstract}
\vspace{2 mm}
\begin{quote}

	Renormalization group (RG) and resummation techniques have been used in $N$-component $\phi^4$ theories at fixed dimensions below four to determine the presence of non-trivial IR fixed points and
	to compute the associated critical properties. Since the coupling constant is relevant in $d<4$ dimensions, the RG is entirely governed by renormalization scheme-dependent terms.
	We show that the known proofs of the Borel summability of observables depend on the renormalization scheme and apply only in  ``minimal"  ones, equivalent in $d=2$ to an operatorial normal ordering
	prescription, where the $\beta$-function is trivial to all orders in perturbation theory.
	The presence of a non-trivial fixed point can be unambiguously established by considering a physical observable, like the mass gap, with no need of RG techniques.
	Focusing on the $N=1$, $d=2$ $\phi^4$ theory, we define a one-parameter family of renormalization schemes where Borel summability is guaranteed and study the accuracy on the determination of the
	critical exponent $\nu$ as the scheme is varied. While the critical coupling shows a significant sensitivity on the scheme, the accuracy in $\nu$ is essentially constant.
	As by-product of our analysis, we improve the determination of $\nu$ obtained with RG methods by computing three more orders in perturbation theory.

\end{quote}

\newpage

\tableofcontents

\section{Introduction}

Perturbative Renormalization group (RG) techniques have been extensively used in the analysis of critical phenomena that have a continuum limit
description in terms of an effective Landau-Ginzburg Hamiltonian. A notable class of models of this kind is represented by $N$-component $\phi^4$ theories in $2\leq d <  4$ dimensions (see ref.\cite{Pelissetto:2000ek} for a review that includes an extensive list of references). The two main approaches developed in the past use RG methods either starting from $d=4-\epsilon$ dimensions, the so called
$\epsilon$-expansion \cite{Wilson:1971dc}, or by keeping the space dimension fixed at $d=2,3$, respectively \cite{Parisi:1993sp}.
In the $\epsilon$-expansion we can establish in a renormalization scheme-independent and analytic way the existence of weakly coupled fixed points for $\epsilon\ll1$.
Moreover, we can directly study the critical theory because of the absence of IR divergences.
At fixed dimension, the latter forbids a direct study of the critical regime. One instead considers the massive theory and establishes the presence of IR fixed points by
looking for zeroes of a properly defined $\beta$-function. These are necessarily strongly coupled and require a Borel resummation of the perturbative series to be established.
Borel resummation is needed also in the $\epsilon$-expansion if one wants to reach physical dimensions at $\epsilon=1$ or at $\epsilon=2$, given the asymptotic nature of the associated series.

The Borel summability of the fixed dimension perturbative series in the $\phi^4$ theory at parametrically small coupling and in the unbroken phase $m^2>0$ has been established long ago \cite{Eckmann,Magnen}
(see also ref.\cite{Eckmann:1979pr}) and recently extended for finite values of the coupling and to a more general class of scalar field theories, including for instance $d=3$ $N$-component $\phi^4$ theories with both $m^2>0$ and $m^2<0$, using steepest descent arguments \cite{Serone:2018gjo,Serone:2019szm}. The Borel summability of the $\epsilon$-expansion in the $\phi^4$ theory remains instead to be proven.\footnote{The main difficulty here is to find a non-perturbative definition of the theory in $d$-dimensions for non-integer $d$. So far the Borel summability is assumed based on the large-order behavior of the series \cite{Brezin:1976vw} and the successful results of numerical resummations at finite order, see e.g. ref.\cite{Kompaniets:2017yct} for a recent six-loop study. Some of the arguments used in ref.\cite{Serone:2018gjo} to prove Borel summability formally apply also for the $\epsilon$-expansion, providing somehow further evidence in support of it.}

The aim of this paper is to study the relation between the RG flow, the renormalization scheme (RS) dependence and the Borel summability in $N$-component $\phi^4$ theories.
We will only consider perturbative series at fixed dimensionality, given the lack of a proof of Borel summability for the $\epsilon$-expansion.
For simplicity, we will focus on $O(N)$ vector models (in the unbroken phase), though generalizations to other models, such as theories with cubic anisotropy, should not present any difficulty.
The close connection between RG and RS is evident if one recalls that in $d<4$ the quartic coupling is a relevant parameter. As such, there are no RS-independent logs to be resummed and
the $\beta$-function of the theory is entirely given by RS-dependent terms. Ironically,  the results of both the early works \cite{Eckmann,Magnen} and the recent ones \cite{Serone:2018gjo,Serone:2019szm}
are based on a ``minimal" RS, equivalent to an operatorial normal ordering prescription in $d=2$, where the $\beta$-function is trivial to all orders in perturbation theory!

We start in section \ref{sec:LL4D} by reviewing known facts about RS-dependence of relevant (and irrelevant) couplings. In subsection \ref{subsec:dl4} we then review the approach proposed in ref.\cite{Parisi:1993sp}, based on a properly defined $\beta$-function, and recall in subsection \ref{subsec:noRG} how the critical regime can be analyzed in minimal RSs (already used in ref.\cite{Serone:2018gjo} for the study of the $d=2$ $\phi^4$ theory) with no need of RG  techniques.

In section 3 we briefly review the proofs of Borel summability of scalar field theories  of refs.\cite{Eckmann,Magnen} and ref.\cite{Serone:2018gjo}, emphasizing their RS-dependence.
In particular, we show that the arguments of ref.\cite{Serone:2018gjo} do not straightforwardly apply in RSs such as the one used in ref.\cite{Parisi:1993sp}  based on momentum subtraction, essentially because they lead to counterterms that have an infinite number of contributions in perturbation theory and involve quartic $\phi^4$ terms that dominate the path integral for large field configurations. This should be contrasted with the minimal RSs, where only one mass counterterm is introduced, that can be determined in closed form in perturbation theory, having only a finite number of terms, and does not change the convergence property of the classical Hamiltonian. It should be stressed that we are not claiming here that the theory is not Borel resummable in RSs such as the one in ref.\cite{Parisi:1993sp}, 
but that this does not automatically follow from proofs performed in other RSs \cite{Eckmann,Magnen,Serone:2018gjo}.

In section 4 we focus on the $d=2$  ($N=1$) $\phi^4$ theory and perform a more quantitative study on the RS-dependence of the Borel summability of the Schwinger two-point function.
We define a one-parameter family of RSs as normal ordering with respect to a parameter $\mu\neq m$
where Borel summability is guaranteed. We then analyze, at fixed number of terms kept, the sensitivity on the scheme of RS-dependent quantities such as the mass gap $M$ as a function of the coupling and the critical coupling $g_c$.
Perhaps more importantly, we study the effectiveness of the Borel resummation as the RS is varied, by computing the accuracy in the determination of a physical observable
(the critical exponent $\nu$). As expected, the numerical values of the computed first-order terms of the perturbative series significantly depend on the RS and more or less differ from
the expected large-order behavior. On the other hand, when perturbation theory remains applicable as the RS is varied,  the accuracy in $\nu$ remains essentially constant and like the one
determined in the standard normal ordering scheme with $\mu=m$ \cite{Serone:2018gjo}.
We conclude in section 5. We report in appendix \ref{sec:appendix} an improvement in the determination of $\nu$ obtained by RG methods in the RS of ref.\cite{Parisi:1993sp} by adding three more orders to the known perturbative series and numerically resumming the resulting series.

\section{RG and RS Dependence}

\label{sec:LL4D}

It is well-known that RG techniques allow us to improve the perturbative expansion by resumming certain logarithmic (leading, next to leading, etc) contributions \cite{GellMann:1954fq}.
Before discussing the RS dependences of RG flows and possibly of the Borel summability,
it is useful to review basic known facts about the RS dependence of $\beta$-functions  and the uses of RG in $\phi^4$ theories in $d<4$.
There are no new results in this section, so the expert reader might skip it and pass directly to section 3.

By definition, the $\beta$-function coefficients in mass-independent RSs (such as minimal subtraction when using dimensional regularization)
depends only on the coupling constants and not on ratios of the sliding scale $\mu$ with mass parameters.
In particular, in presence of $p$ marginal couplings $g_i$ ($i=1,\ldots, p)$, we have
\be
\beta_i = \mu \frac{dg_i}{d\mu}  = \beta^{ijk}_0 g_j g_k + {\cal O}(g^3),
\ee
where $\beta^{ijk}_0$ are constants.
Among mass-independent RSs, the leading $\beta$-function coefficients $\beta^{ijk}_0$ are RS-independent (for $p=1$ the next to leading term is also RS-independent).
More in general, in presence of dimensionful  couplings,  the $\beta$-function coefficients can be RS-dependent, even when mass effects are neglected.
Indeed, if we denote by $g_i$ and $\widetilde g_i$ the couplings in two different RSs,
we have by dimensional analysis
\be
\widetilde g_i = g_i +  \mu^{\Delta_i - \Delta_j-\Delta_k} c_{ijk} g_j g_k +{\cal O}(g^3)\,,
\ee
where $\Delta_i$ denote the classical scaling dimensions of the couplings $g_i$ and $c_{ijk}$ are constant coefficients. Dimensional analysis fixes also
the form of the perturbative expansion of the $\beta$-functions. In terms of the dimensionless rescaled couplings $h_i = \mu^{-\Delta_i} g_i$'s, we have
\be
\beta_i = \mu\frac{dh_i}{d\mu}  = -\Delta_i h_i + b^{ijk}_0 h_j h_k + {\cal O}(h^3) \,,
\ee
where $b^{ijk}_{0}$ are constant coefficients.
It is immediate to get the relation between the two leading $\beta$-function coefficients in the two RSs:
\be
\widetilde b^{ijk}_0 =  b^{ijk}_0  + c_{ijk} (\Delta_i - \Delta_j-\Delta_k)\,.
\ee
Universal coefficients arise when $\Delta_i - \Delta_j-\Delta_k = 0$. Renormalization schemes where classical dimensional analysis is preserved at the quantum level, like minimal subtraction when using dimensional regularization, give automatically $b^{ijk}_0=0$ when $\Delta_i - \Delta_j-\Delta_k \neq 0$, and only keep the RS-independent coefficients.\footnote{This property is a consequence of dimensional regularization of setting to zero power-like divergences and keeping only the logarithmic ones. Logarithmic divergences are the only ones not saturated by UV physics and sample uniformly all energy scales. Since the IR physics should be insensitive to the details of the different RSs, we conclude that the associated $\beta$-function coefficients should be RS-independent. In presence of dimensionful couplings dimensional regularization is no longer a mass-independent RS, since by dimensional analysis $\beta$-functions can depend on masses.} Such dimensional arguments have important simple implications, not always fully appreciated in the literature. For instance, in an effective field theory with irrelevant couplings and no relevant or marginal couplings, in the limit where mass effects are negligible,
the irrelevant coupling $\lambda$ with the smallest dimension $\Delta$ has a trivial $\beta$-function  to {\it all orders in perturbation theory}:
\be
\beta_{\lambda}= - \Delta \lambda\,,  \quad  (\lambda\;\; {\rm smallest \;\; irrelevant \;\; coupling)}\,.
\label{eq:betadg4}
\ee
This is the case, for instance, for the gauge coupling in Yang-Mills or for the quartic coupling of a $\phi^4$ effective theory in $d\geq 5$. In these cases
there is no analogue of the log resummation needed in treating marginal couplings and hence no need to improve the perturbative expansion.
On the other hand, in mass-dependent RSs, such as momentum subtraction, dimensional arguments do not apply and in general $\beta_{\lambda}$ is non-trivial.
The corresponding RG flow that one obtains amounts to resum RS-dependent threshold effects.

\subsection{Use of RG Flows in \texorpdfstring{$\phi^4$}{phi**4}  Theories in $d<4$: the RS \texorpdfstring{$\widetilde {\cal S}$}{Stilde} }
\label{subsec:dl4}

The bare Landau-Ginzburg Hamiltonian (or equivalently the euclidean action) for $O(N)$-invariant $\phi^4$ vector models reads
\be
H_0 = \int d^d x \Bigl[\frac12 (\partial_\mu \vec \phi_0)^2+\frac 12 m_0^2 \vec\phi_{0}^2 + \lambda_0 (\vec \phi_0 ^2)^2  \Bigr]\,,
\label{eq:H0}
\ee
where $\vec \phi_0$ is an $N$-component scalar. In $d=2$ and $d=3$, scalar theories with up to quartic couplings are super-renormalizable. Aside from the ground state energy, which will be neglected from now on, only the mass term requires renormalization, the coupling constant $\lambda_0$ and the elementary field (wave function renormalization) being finite.
Finiteness of the coupling immediately implies that in a mass-independent RS (e.g. minimal subtraction) we have a trivial $\beta$-function
to all orders in perturbation theory. There are no logs to be resummed and, like in the $d=5$ case mentioned below eq.~(\ref{eq:betadg4}), no RS-independent terms appear in $\beta$, besides the classical contribution.\footnote{It is important to emphasize here that the minimal subtraction scheme alluded before is different from the so called ``minimal subtraction without $\epsilon$-expansion" introduced in refs.\cite{Schloms:1989rq,Schloms:1990zz} and sometimes used in the statistical physics community. By minimal subtraction we mean the renormalization procedure with minimum impact on the Hamiltonian, where counterterms are introduced only when necessary to cancel divergences in the actual dimensionality one is considering. It is not necessarily related to dimensional regularization.
	In a super-renormalizable theory, such counterterms contain a finite number of terms in perturbation theory.
	In contrast, in the ``minimal subtraction without $\epsilon$-expansion" of refs.\cite{Schloms:1989rq,Schloms:1990zz}, counterterms are computed using  minimal subtraction with dimensional regularization in $d=4$, and as such contain an infinite number of terms in perturbation theory.} Yet, one can define a non-minimal RS where $\beta$ is non-trivial and study the associated RG flow  \cite{Parisi:1993sp}.
Although the wave function renormalization $Z$ and the coupling constant counterterm $Z_{\widetilde g}$ are not necessary, yet we can define renormalized parameters using a momentum subtraction RS like
we would do in $d=4$ as follows:
\be
\vec\phi_{0} = \sqrt{Z} \vec\phi\,, \quad Z m_0^2 =  \widetilde M^2  +\delta  \widetilde M^2\,, \quad \lambda_0 =  \widetilde M^{4-d}  \widetilde g \frac{Z_{ \widetilde g}}{Z^2} \,,
\label{RenConddl4}
\ee
and fix the counterterms by the following three conditions at zero momentum:
\be\begin{split}
	\Gamma^{(2)}_{ij}(p=0) & = \delta_{ij} \widetilde M^2\,, \quad \frac{d\Gamma_{ij}^{(2)}(p=0)}{dp^2} =\delta_{ij}\,, \\
	\Gamma^{(4)}_{ijkl}(p_1=p_2=p_3 = 0) & =  8 \widetilde g \,  \widetilde M^{4-d}   (\delta_{ij} \delta_{kl} + \delta_{ik} \delta_{jl} + \delta_{il} \delta_{jk} ) \,,
	\label{RenCond}
\end{split}
\ee
where $i,j,k,l$ are $O(N)$ indices and as usual $\Gamma^{(n)}$ are related to the bare 1PI Schwinger functions $\Gamma_0^{(n)} $ as $\Gamma^{(n)}=\Gamma_0^{(n)} Z^{n/2}$.
We will denote the RS defined by eq.~(\ref{RenCond}) as $\widetilde {\cal S}$. Correspondingly all parameters in the RS $\widetilde {\cal S}$ will be labeled with a tilde.
No sliding scale $\mu$ needs to be introduced, yet a $\beta$-function can be defined in the spirit of the original Callan and Symanzik derivation  \cite{Callan:1970yg,Symanzik:1970rt}.
Since $\lambda_0$ does not depend on the physical mass $ \widetilde M$, we have $ \widetilde M\, d\lambda_0/d \widetilde M=0$, that gives rise to the following equation
\be
\widetilde \beta(\widetilde g) \equiv   \widetilde M \frac{d  \widetilde g}{d \widetilde M}\bigg|_{\lambda_0}= (d-4) \Big(\frac{d\log ( \widetilde g Z_{ \widetilde g}/Z^2)}{d \widetilde g}\Big)^{-1} \,.
\label{betaphi24}
\ee
In contrast to the more familiar form of the Callan Symanzik equations for which one has $\mu\, d\Gamma_0^{(n)}/d\mu=0$ and thus homogeneous equations in $\Gamma^{(n)}$, we  now have (omitting $O(N)$ indices)
\be
\left[ \widetilde M\frac{\partial}{\partial  \widetilde M} +  \widetilde \beta( \widetilde g)\frac{\partial}{\partial  \widetilde g} -\frac{n}{2}  \widetilde \eta( \widetilde g) \right] \Gamma^{(n)} =  \widetilde M^2
\widetilde \sigma  \Gamma^{(n,1)} \,,
\label{eq:CSeqs}
\ee
where $\Gamma^{(n,1)}$ are the 1PI $n$-point functions with one insertion of the renormalized composite operator $\phi^2=\phi_0^2 Z_{\phi^2}$ at zero momentum, while the parameters $\widetilde \eta$ and $\widetilde \sigma$ are defined as follows:
\bea
\widetilde \eta( \widetilde g) & = &  \widetilde M\frac{d \log Z}{d  \widetilde M}\bigg|_{\lambda_0} = \widetilde  \beta( \widetilde g) \frac{d \log Z}{d  \widetilde g}\,,  \label{eq:anomDim} \\
\widetilde \sigma &= & \frac{1}{Z_{\phi^2}} \frac{1}{ \widetilde M^2}   \widetilde M\frac{d m_0^2}{d  \widetilde M}\bigg|_{\lambda_0}\,.
\eea
The counterterm $Z_{\phi^2}$ can be fixed by demanding
\be
\Gamma^{(2,1)}_{ij}(p_1=p_2=0) = \delta_{ij}\,.
\label{Gamma21Def}
\ee
From $Z_{\phi^2}$ and $Z$ we can determine the anomalous dimension of $\phi^2$ as
\be
\widetilde \eta_{\phi^2} = -  \widetilde M\frac{d \log (Z_{\phi^2}/Z)}{d  \widetilde M}\bigg|_{\lambda_0} = - \widetilde \beta( \widetilde g) \frac{d \log (Z_{\phi^2}/Z)}{d  \widetilde g}\,.
\label{eq:anomDimphi2}
\ee
Consistency between eq.~(\ref{RenCond}) and the CS equation (\ref{eq:CSeqs}) with $n=2$ gives
\be
\widetilde \sigma = 2 - \widetilde  \eta\,.
\ee
In $d=2$, $N=1$ and $d=3$ for any $N$, starting from the unbroken phase ($m^2>0$), we expect there should exist a critical
value of the coupling $\widetilde g$ where the theory becomes gapless and a CFT arises.\footnote{Non-unitary CFTs can arise for other values of $N$, such as $N=0$, which describes self-avoiding random walks.}${}^{,}$\footnote{The possibility that the critical theory is scale-, but not conformal, invariant has been recently excluded for the $d=3$ $\phi^4$ theory \cite{Meneses:2018xpu}.}
This non-trivial IR fixed point should be visible as a non-trivial zero of $\widetilde \beta$. The expansion of $\widetilde \beta$ in perturbation theory reads
\be
\widetilde \beta = (d-4) \widetilde g +{\cal O}(\widetilde g^2) \,.
\ee
When $d=3$ or $d=2$, a non-trivial zero is necessarily strongly coupled. The presence of a non-trivial fixed point cannot be established perturbatively,
but it can at the non-perturbative level, namely upon Borel resumming the perturbative expansion.
Several resummation of the $\beta$-function $\widetilde \beta(\widetilde g)$ over the years have shown indeed
the presence of a zero for some non-perturbative value of the coupling in $d=2$ and $d=3$
\cite{Baker:1977hp,LeGuillou:1979ixc,Orlov:2000wn,Sokolov:2005xp,Guida:1998bx,Serone:2018gjo}.
In the RS $\widetilde {\cal S}$ this zero defines the critical coupling $\widetilde g_c$:
\be
\widetilde \beta( \widetilde g_c) = 0 \,.
\ee
When we approach the critical regime the correlation length diverges, $\widetilde M\rightarrow 0$, the right-hand side in eq.~(\ref{eq:CSeqs}) can be neglected
and the Schwinger 1PI functions $\Gamma^{(n)}$ satisfy the scaling relations valid for a conformal invariant  theory.
Once $ \widetilde g_c$ is determined, one can Borel resum the perturbative series in $ \widetilde g$ for $ \widetilde \eta$ and $ \widetilde \eta_{\phi^2}$ and identify the (RS independent) critical exponents $\eta$ and $\nu$ as
\be
\eta  \equiv  \widetilde \eta( \widetilde g_c) \,, \quad \eta_{\phi^2}  \equiv \widetilde \eta_{\phi^2}( \widetilde g_c)   \,, \quad \nu =\frac{1}{ 2-\eta_{\phi^2}} \,.
\label{eq:etanuRG}
\ee
Other critical exponents can be obtained using scaling relations.

\subsection{No Use of RG Flows in \texorpdfstring{$\phi^4$}{phi**4} Theories in $d<4$: the RS ${\cal S}$}
\label{subsec:noRG}

As we have discussed in the previous subsection, only the mass term in the Hamiltonian (\ref{eq:H0}) requires renormalization, up to 1- and 2-loops in $d=2$ and $d=3$, respectively.
Correspondingly, we can directly equate bare fields and quartic couplings with their renormalized  counterparts and introduce only a mass counterterm $\delta m^2$:
\be
\vec\phi_0 = \vec\phi\,, \quad m_0^2 = m^2 + \delta m^2\,, \quad \lambda_0 = \lambda \,.
\label{eq:RSNorm}
\ee
We define the RS ${\cal S}$ \footnote{A RS similar to ${\cal S}$ has been introduced for $d=3$ $O(N)$ models
	in ref.\cite{Guida:2005bc}, where it was dubbed $I$ and was meant to be an intermediate step towards the final RS $\widetilde {\cal S}$ (dubbed $M$ in that paper).} by imposing the following conditions for the mass counterterm $\delta m^2$:
\begin{equation}
	\begin{aligned}
		\raisebox{-.42\height}{\includegraphics[width=1.3cm]{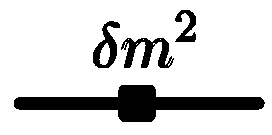}} + \raisebox{-.33\height}{\includegraphics[width=1.7cm]{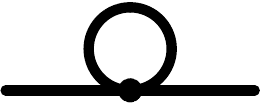}}  = 0  \quad & (d=2)\,,  \\
		\raisebox{-.42\height}{\includegraphics[width=1.3cm]{img/deltam.png}} + \raisebox{-.33\height}{\includegraphics[width=1.7cm]{img/2pt-1t.pdf}} +
		\raisebox{-.63\height}{\includegraphics[width=2.1cm]{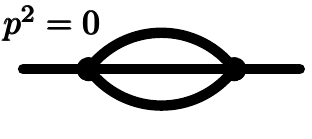}} = 0 \quad                                                                         & (d=3) \,,
		\label{eq:MSm}
	\end{aligned}
\end{equation}
To all orders in perturbation theory we have
\begin{equation}
	\frac{\delta m^2}{m^2} =  a^{(d)}_{1} g + \delta_{d,3} \, a^{(d)}_{2} g^2 \,,
\end{equation}
where $a^{(d)}_{i}$ are divergent coefficients and $g$ is the effective dimensionless coupling constant defined as
\be
g = \frac{\lambda}{m^{4-d}}\,.
\ee
Callan-Symanzik equations like eq.~(\ref{eq:CSeqs}) can be considered also in the RS ${\cal S}$. In this case one simply gets
\be
\left[ m\frac{\partial}{\partial m} +  \beta(g)\frac{\partial}{\partial  g} -\frac{n}{2}  \eta(g) \right] \Gamma^{(n)}_0 =  m^2  \sigma  \Gamma^{(n,1)}_0 \,,
\label{eq:CSeqsRSs}
\ee
where
\be
\beta(g) \equiv m \frac{dg}{dm}\Big|_{\lambda_0}  = (d-4) g\,,
\qquad
\eta = 0\,,
\qquad
\sigma = 2  \left \{
\begin{array}{lc}
	1+ \frac{N+2}{\pi} g\,, \quad                          & d=2\,,\vspace{8pt} \\
	1 + \frac{N+2}{2\pi} g -2 \frac{(N+2)}{\pi^2} g^2\,, ~ & d=3\,.
\end{array} \right.
\label{CSMS2}
\ee
Note that the mass $m$ entering eq.~(\ref{eq:CSeqsRSs}) is the renormalized mass $m$ and {\it not} the pole mass $M$ as defined in eq.~(\ref{mphDef}) below.
As a consequence, in the critical regime $M\rightarrow 0$ the term proportional to $\Gamma^{(2,1)}_0$ does not vanish, in contrast to what happens in eq.~(\ref{eq:CSeqs}) in the RS $\widetilde {\cal S}$ when
$\widetilde M\rightarrow 0$. Hence demanding $\beta=0$ in eq.~(\ref{eq:CSeqsRSs}) does not correspond to a (non-trivial) critical regime and no interesting RG flow is expected
from $\beta$ in eq.~(\ref{CSMS2}). The latter equations can however be useful, as we will show in the appendix.

Renormalization group methods are not essential and we can access the critical regime by a direct computation of observables.
One can define the pole mass $M$ as the zero of $\Gamma^{(2)}_0$
for complex values of the Euclidean momentum:
\be
\Gamma^{(2)}_{0,ij}(p^2=-M^2)\equiv 0\,.
\label{mphDef}
\ee
The critical coupling can be determined directly as the value of $g$ where the theory is gapless:
\be
M(g_c) = 0\,.
\label{gcDef}
\ee
The critical exponent $\nu$, defined as
\be
M(g) \propto |g_c-g|^\nu , \quad g\rightarrow g_c \,,
\label{nuDef}
\ee
can be computed by resumming a properly defined function of $M^2$. For instance, if
\be
L(g)\equiv \frac{2g^2}{g \partial_g \log M^2}\,,
\label{eq:Lg}
\ee
$\nu$ can be extracted as \cite{Serone:2018gjo}
\be
\nu = \frac{g_c}{\partial_g L}\bigg|_{g=g_c}\,.
\label{nuextr}
\ee
The exponent $\eta$ can be determined directly from its definition as the power-like decay of the two-point function at the critical point:
\be
\langle \phi_i(x) \phi_j(0) \rangle_{g=g_c} \approx \frac{\delta_{i,j}c_{\phi}}{|x|^{d-2+\eta}} \,, \quad i,j=1,\ldots, N\,,
\label{etaDef}
\ee
where $c_\phi$ is a constant. This is the approach that has been taken in ref.\cite{Serone:2018gjo} to determine $g_c$, $\nu$ and $\eta$ in the $d=2$ $\phi^4$ theory.
Alternatively, $\nu$ could be determined more directly by means of eq.~(\ref{eq:etanuRG}), where  $\eta_{\phi^2}$
is extracted as
\be
\langle \phi^2(x) \phi^2(0) \rangle_{g=g_c} \approx \frac{c_{\phi^2}}{|x|^{2(d-2+\eta_{\phi^2})}} \,,
\label{etaphi2Def}
\ee
where again $c_{\phi^2}$ is a constant. It is worth emphasizing that the value of $\eta$ found using the above procedure in the RS ${\cal S}$ is in good agreement with the exact result $\eta=1/4$, while a long standing mismatch is found when using eq.~(\ref{eq:etanuRG}) in the RS $\widetilde {\cal S}$. A similar long standing mismatch occurs in the evaluation of $\omega\equiv \widetilde \beta'(\widetilde g_c)$ in the RS $\widetilde {\cal S}$, which significantly differs from the exact value $\omega=2$. We have verified that no improvement is achieved by resumming $\omega$ from the expression of $\widetilde\beta$
reported in eq.~(\ref{eq:A2}) and derived in ref.\cite{Serone:2018gjo}, which includes two more orders in the known perturbative expansion.
These problems seem to be related to possible non-analyticities in $ \widetilde \beta( \widetilde g)$ that give rise to a poor convergence of the numerical Borel resummation to the exact result \cite{Pelissetto:1997gk,Calabrese:2000dy}, though they might also be a signal of absence of Borel summablity in the RS $\widetilde {\cal S}$ for such observables.
It would be interesting to check if the mismatch for $\omega$ disappears (like for $\eta$) if the RS ${\cal S}$ is used and $\omega$
extracted directly from a two-point function, such as $\langle \phi^4(x) \phi^4(0) \rangle_{g=g_c}$.\footnote{We thank A. Pelissetto for drawing our attention to the critical exponent $\omega$.}

A direct approach to the critical regime without the use of RG techniques allows us  to bypass the need of evaluating the 4-pt function $\Gamma^{(4)}$. The number of diagrams with $L$ loops in a $2n$-pt function
$\Gamma^{(2n)}$ is expected to scale as the number of loop diagrams in the vacuum energy with $L+n$ loops.
This is seen by noting that if we connect the $2n$ external lines in pairs, we get a vacuum energy graph with $n$ more loops. Large order estimates confirm this expectation.
At fixed number of loops $L$, then, evaluating the 4-pt function is computationally more challenging than evaluating the 2-pt function, due to the larger number of Feynman diagrams.
In $d=2$, the RS ${\cal S}$ is equivalent, in an operatorial formalism, to normal order the operators with respect to the mass $m$ and
has been used in the literature as a reference RS to compare various non-perturbative computations of the critical coupling \cite{Milsted:2013rxa,Rychkov:2014eea,Bosetti:2015lsa,Pelissetto:2015yha,Elias-Miro:2017xxf,Elias-Miro:2017tup,Bronzin:2018tqz,Kadoh:2018tis}.

Note that the definition of $g_c$ given by eq.~(\ref{gcDef}) could be adopted also in the RS $\widetilde {\cal S}$, bypassing the evaluation of the beta-function $\widetilde \beta$. Similarly one could compute
$\eta$ and $\eta_{\phi^2}$ directly from eqs.~(\ref{etaDef}) and (\ref{etaphi2Def}).

\section{Borel Summability and RS Dependence}
\label{BorelScheme}

In this section we review previous results on the Borel summability of Schwinger functions in the $\lambda \phi^4$ theory, contained in both the early
\cite{Eckmann,Magnen} and the more recent papers \cite{Serone:2018gjo,Serone:2017nmd}, and show how they depend on the RS.

\begin{figure}[t!]
	\includegraphics[width=0.34\textwidth]{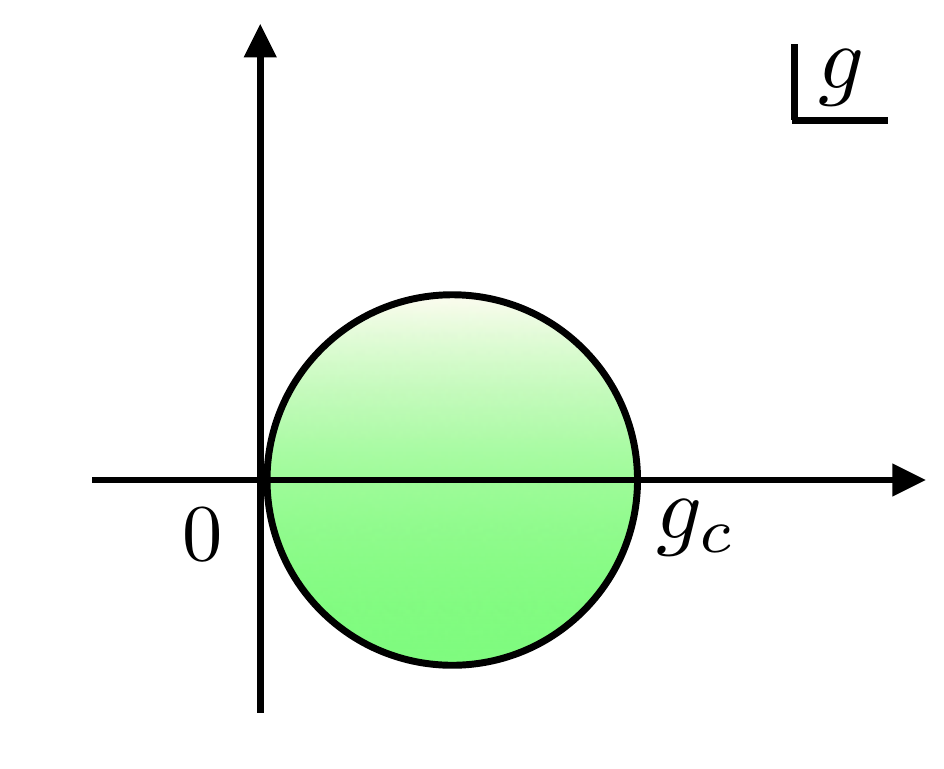}
	\centering
	\caption{The green disk shows the minimum region of analyticity for a Borel resummable function \cite{Sokal:1980ey} (the origin is excluded).}
	\label{fig:Sokal}
\end{figure}

We first briefly review the early proofs of the Borel summability of Schwinger functions in the $d=2$ \cite{Eckmann} and $d=3$ \cite{Magnen} $N=1$ $\phi^4$ theory.
These papers are in the context of constructive quantum field theory,  an area of research particularly active in the late 60s and in the 70s, that tries to give a rigorous mathematical foundation to quantum field theories, see e.g. ref.\cite{Jaffe:2000ub} for an overview. We do not enter into details, but only mention the key steps and the logic followed in these papers, focusing on the RS chosen.
As starting point the bare Hamiltonian (\ref{eq:H0}) is renormalized by adding mass counterterms (we neglect vacuum energy counterterms) as in eq.~(\ref{eq:RSNorm}). In particular, bare and
renormalized fields and couplings are identified. In $d=2$ the RS chosen in ref.\cite{Eckmann} is identical to normal ordering with respect to the mass $m^2$ and hence coincides with the RS ${\cal S}$.
The renormalization conditions in $d=3$ are not given in an explicit form in ref.\cite{Magnen},
but they are essentially equivalent to the RS ${\cal S}$. In particular, only ${\cal O}(\lambda)$ and $O(\lambda^2)$ mass counterterms are present,\footnote{In eq.(2.1.1) of ref.\cite{Magnen} only the $O(\lambda^2)$ term appears, the $O(\lambda)$ one being hidden in the normal ordering operation.} as in the condition (\ref{eq:MSm}) defining the RS ${\cal S}$.
In both $d=2$ and $d=3$, the local operators $\phi(x_i)$ are smeared with sufficiently regular functions $f_i$ with compact support around a region surrounding $x_i$ to define a field $\phi_{f_i} = \int d^2x_i f(x_i) \phi(x_i)$.
Finally, it is shown that for $|\lambda|<\epsilon$, ${\rm Re} \,\lambda>0$ and large enough $m^2>0$ (i.e. at parametrically weak coupling $g\ll 1$ in our notation), the
$2n$-point smeared Schwinger functions
\be
G^{\rm sm}_{2n}(\lambda) = \frac{  \int {\cal D}\phi\, \phi_{f_1}\ldots \phi_{f_{2n}}\, e^{-H_0[\phi]} }{  \int {\cal D}\phi\, e^{-H_0[\phi]} }
\label{eq:anaGsm}
\ee
are analytic in $\lambda$ with bounded derivatives:
\be
\left|  \frac{d^k}{d\lambda^k} G^{{\rm sm}}_{2n}(\lambda) \right|  \leq C_1 C_2^k k!^2\,,
\label{eq:boundG}
\ee
with $C_1$ and $C_2$ two constants.\footnote{The bound (\ref{eq:boundG}) found in ref.\cite{Magnen} is actually proportional to $k!^{2+\xi}$, with $\xi>0$, and one has to generalize Watson criterion to show Borel summability. We thank J. Magnen for discussions on this issue and for pointing out that the limit $\xi\rightarrow 0$ might be taken by using the so called multiscale expansion \cite{Rivasseau}.}  Under suitable conditions on the smearing functions $f_i$, the analyticity domain of $G^{\rm sm}_{2n}(\lambda)$ can be extended to a region including points where
${\rm Re}\, \lambda<0$. The asymptotic series of the smeared Schwinger functions $G^{\rm sm}_{2n}(\lambda)$ satisfy then the sufficient criterion for Borel summability as given by Watson (see e.g. theorem 136 in chap.VIII of ref.\cite{Hardy}). Soon after, it was pointed out that the analytic continuation to a region including points where ${\rm Re}\, \lambda<0$ is unnecessary.
One can instead use a necessary and sufficient criterion of Borel summability, found long ago by Nevanlinna and rediscovered in ref.\cite{Sokal:1980ey}, that requires a domain of analyticity only in a region with ${\rm Re} \,\lambda>0$, see fig.~\ref{fig:Sokal}. In the $d=2$ case \cite{Eckmann} the analyticity of the Schwinger functions is extended for more general functions involving normal-ordered composite operators of the form $\phi^q$, with $q$ a positive integer, and for generic bounded polynomial potentials with degree $P$. In this case, the factor $k!^2$ in eq.~(\ref{eq:boundG}) is replaced by $k!^{P/2}$.\footnote{Note that the ordinary Watson criterion for Borel summability requires
	$P=4$. Presumably this is the reason why the authors \cite{Eckmann} did not discuss Borel summability of theories with higher order interaction terms. On the other hand, the arguments made in ref.\cite{Serone:2018gjo} and reviewed in what follows allow us to conclude that these theories are Borel resummable in the proper loopwise expansion. For instance, for a $\phi^{2p}$ potential the loopwise parameter is $g_{2p} = (\lambda/m^2)^{1/(p-1)}$ and Schwinger functions are Borel summable in $g_{2p}$, though they are not in $\lambda/m^2$.}

We now review and expand a bit some of the considerations made in ref.\cite{Serone:2018gjo} about the Borel summability of scalar field theories in $d<4$.
For concreteness we focus on $O(N)$ vector models, though most considerations apply more in general.
Consider a $2n$-point Schwinger function
\begin{equation}
	G^0_{2n}(x_1,\ldots,x_n)={\cal N}  \int {\cal D}\phi_0\, \phi_0(x_1)\ldots \phi_0(x_{2n})\, e^{-H_0[\phi_0]} \,,
	\label{eq:G0n}
\end{equation}
where ${\cal N}$ is an irrelevant constant factor, we omitted $O(N)$ indices,  and $H_0$ is the bare Hamiltonian (\ref{eq:H0}). %
We renormalize the theory in the RS ${\cal S}$ using eqs.~(\ref{eq:RSNorm}) and (\ref{eq:MSm}).
It is useful to rescale fields and coordinates as follows:
\be
\vec \phi(x) = \frac{m}{\sqrt{\lambda}} \vec \Phi(y)\,, \quad y = mx\,,
\label{eq:chFi}
\ee
and rewrite eq.~(\ref{eq:G0n}), omitting also the space dependences, as
\be
G_{2n}=G^0_{2n}={\cal N}' g^{-n} m^{(d-2)n} \int {\cal D}\Phi\, \Phi(y_1)\ldots \Phi(y_{2n})\, e^{-\delta H[\Phi]} e^{-H[\Phi]/g} \,, \quad g = \frac{\lambda}{m^{4-d}}\,,
\ee
with
\bea
H[\Phi] & = & \int d^dy\, \Bigl[\frac12 (\partial \vec \Phi)^2+\frac 12 \vec \Phi^2 +  (\vec \Phi^2)^2  \Bigr]\,, \label{Sren} \\
\delta H[\Phi] & = &
\left( a^{(d)}_{1} + \delta_{d,3} \, a^{(d)}_{2} g \right) \int d^dy\, \Bigl[ \frac 12 \vec \Phi^2 \Bigr]\,.
\label{SrenCTI}
\eea
The counterterm Hamiltonian $\delta H$, in both $d=2$ and $d=3$, is subleading to $H$ in a saddle point expansion in $g$  and does not change the saddle point structure of
$H[\Phi]$ if the convergence of the path integral at large field values is dictated by $H$. This is the case in the RS ${\cal S}$, since $\delta H$ is quadratic in the field, while $H$ is quartic.
A simple scaling argument, equivalent to an euclidean version of Derrick's theorem \cite{Derrick:1964ww,Brezin:1976wa}, allows us to show that the Hamiltonian  $H$ does not have any non-trivial critical points with finite energy, aside from the trivial one $\vec\Phi=0$,  for real field configurations. The combination of reality and boundedness of the Hamiltonian and the presence of a unique critical point makes the domain of integration of the path integral (\ref{eq:G0n}) a single Lefschetz thimble, guaranteeing  the Borel summability of the Schwinger functions $G^{(2n)}$ \cite{Serone:2017nmd}.\footnote{More precisely, we mean that
for any choice of non-coincident points $x_i$ the resulting series in $g$ is Borel summable. Alternatively, as in refs.\cite{Eckmann, Magnen}, we could smear the local operators $\vec\phi(x_i)$ by means of some functions $f_i$ and consider their smeared version $\vec\phi_{f_i}$. The asymptotic series of the smeared Schwinger functions $G^{(2n)}_{{\rm sm}}$ %
would then be Borel summable for any sensible choice of smearing functions $f_i$.\label{footnote}} A similar argument is expected to apply for Schwinger functions involving composite operators constructed out of $\vec\phi$ and their derivatives.

It might be useful to compare the results of refs.\cite{Eckmann,Magnen} with those of ref.\cite{Serone:2018gjo}.
While the proof of refs.\cite{Eckmann,Magnen} requires a detailed study of the analytic properties of the exact Schwinger functions in the coupling constant $\lambda$,
the argument based on Lefschetz thimbles in ref.\cite{Serone:2018gjo} makes it possible to avoid such study and to reach the same conclusion in a simpler way.
Borel summability holds for all {real} values of the coupling where the Schwinger functions are well-defined. On physical grounds, we expect this to hold until the theory undergoes  a phase transition,
in which case Schwinger functions or their derivatives can diverge. We can in turn use the necessary and sufficient criterion of Borel summability of ref.\cite{Sokal:1980ey} to establish
that Schwinger functions should be analytic in the region in fig.~\ref{fig:Sokal}. We expect this region to extend until the critical coupling $g_c$ where a second order phase transition (or of any other kind, for more general theories) occurs.\footnote{Schwinger functions analytically continued past a phase transition might still be physically sensible.
	See refs.\cite{Serone:2018gjo,Serone:2019szm} for more details and for some numerical evidence in the $d=2$ $\phi^4$ case.}
Moreover, the simplicity of the arguments in ref.\cite{Serone:2018gjo} immediately allows us to establish Borel summability for more general theories beyond the $\phi^4$ with positive squared mass term, the subject of study in refs.\cite{Eckmann,Magnen}.\footnote{See ref.\cite{Erbin:2019zug} for a very recent paper where Borel summability in certain low dimensional theories is established in the context of
	constructive quantum field theory.} Of course, the more heuristic derivation of ref.\cite{Serone:2018gjo} does not match the standard of mathematical rigor typically requested in constructive quantum field theory.
In particular, as physicists we do {\it assume} that the Schwinger functions (and the theory itself) exist.
In contrast, in constructive field theory the existence of a non-perturbative definition of the theory is generally the first important point to be established, Borel summability (if any) being a by-product.
Interestingly enough, despite the  approaches and the methodologies substantially differ among refs.\cite{Eckmann,Magnen} and \cite{Serone:2018gjo}, in both cases Borel summability is only established in RSs equal or equivalent to the minimal one we denoted by ${\cal S}$.

It is not difficult to show which additional complications can occur in our construction to prove the Borel summability in other non-minimal RSs.
For instance, by repeating the steps from eq.~(\ref{eq:chFi}) to eq.~(\ref{SrenCTI}) in the RS $\widetilde {\cal S}$, we would define
\be
\vec\phi(x) = \frac{\widetilde M^{(d-2)/2}}{\sqrt{\widetilde g}} \vec\Phi(y)\,, \quad y = \widetilde Mx\,,
\label{eq:chFiN}
\ee
and write
\be
G_{2n}=Z^n G^0_{2n}={\cal N}' Z^n \widetilde g^{-n} \widetilde M^{(d-2)n} \int {\cal D}\Phi\, \Phi(y_1)\ldots \Phi(y_{2n})\, e^{-\delta H[\Phi]} e^{-H[\Phi]/\widetilde g} \,, \ee
with $H[\Phi]$ as in eq.~(\ref{Sren}) and
\be
\delta \widetilde H[\Phi]  =   \int d^dx\, \Bigl[\frac12 \frac{Z-1}{\widetilde g} (\partial \vec \Phi)^2 + \frac 12 \frac{\delta \widetilde M^2}{\widetilde g}  \vec \Phi^2  + \frac{Z_{\widetilde g}-1}{\widetilde g} (\vec  \Phi^2)^2\Bigr]\,.
\label{SreCTIIn}
\ee
The counterterm $\delta \widetilde H$ is still subleading to $H$ in a saddle point expansion in $\widetilde g$, but now two subtleties arise.
First, the counterterms $Z$, $Z_g$ and $\delta  \widetilde M^2$ entering $\delta \widetilde H$ are not expressions that can be computed in closed form.
They can only be determined order by order in perturbation theory but the resulting series are in general asymptotic and would require to be resummed.
In other words, the RS $\widetilde {\cal S}$ is intrinsically
perturbative in nature and hence is not suitable to be used to establish a property of a theory that goes beyond perturbation theory, like its Borel resummability.
Second, in contrast to $\delta H$, $\delta \widetilde H$ contains terms quartic in the field $\vec\Phi$, which
could in principle change the convergence properties of the path integral at large field values as dictated by $H$, and possibly invalidate the statement that a resummation of the saddle point expansion
around solutions of $H$ reconstructs the full result.
In light of that, the Borel summability of the expansion in $\widetilde g$ cannot be assessed.
We are not aware of any paper in the constructive quantum field theory literature where the Borel summability of $d=2$ or $d=3$ field theories is established
or even attempted in non-minimal RSs such as ${\cal \widetilde S}$.

A non-perturbative change of RS of the form $g = g(\widetilde g)$ would not affect Borel summability if this mapping preserves the necessary and sufficient conditions for Borel summability,
namely a region of analyticity as in fig.~\ref{fig:Sokal} in the $\widetilde g$ complex plane and a bound on the growth of the coefficients of its asymptotic expansion.
Unfortunately, we typically do not have access to such non-perturbative mapping, and only know it in perturbation theory.
In this case we will have
\be
\label{eq:gvsgtil}
g \sim \widetilde g + \sum_{k=2}^\infty s_k \widetilde g^k\,.
\ee
In general the above series is asymptotic (that's why the $\sim$ sign instead of an equality), as it happens for instance when relating the coupling $g$ in the RS ${\cal S}$ with the coupling $\widetilde g$ in the RS $\widetilde{\cal S}$. We should then first of all face the problem of proving the Borel summability of the series (\ref{eq:gvsgtil}), in general a non-trivial task.
Even if we can somehow prove that the series (\ref{eq:gvsgtil}) is Borel resummable to its exact form $g(\widetilde g)$,
we will still not be able to prove that Borel summability of Schwinger functions in one RS implies that in the other RS.
Indeed, given an observable $F(g)$ which is Borel reconstructed from its asymptotic series $\sum_{k=0}^\infty F_k g^k$ in the RS with coupling $g$, naively plugging eq.~(\ref{eq:gvsgtil}) in the series for $F(g)$ will not in general give rise to a Borel resummable series in $\widetilde g$. Borel summability of the composed series expansion of $F(g(\widetilde g))$ would follow if $\sum_{k=0}^\infty F_k g^k$ were convergent (see e.g. proposition 2.11 of ref.\cite{Iwaki} or section 4.4c of ref.\cite{Costin}) or if both $F(g)$ and $g(\widetilde g)$ satisfy certain analyticity properties close to the origin which are stronger than the ones required for Borel summability \cite{Auberson:1981xx,Auberson:1985}.
Hence, without further assumptions, we would be unable to prove the Borel summability of the observable $F$ in the RS with coupling $\widetilde g$.

\section{RS Dependence and Borel Resummation in the $d=2$ \texorpdfstring{$\phi^4$}{phi**4} Theory}
\label{sec:2dphi4}

In two dimensions the proof of the Borel summability of the $\phi^4$ theory in the minimal RS ${\cal S}$ of ref.\cite{Serone:2018gjo} can trivially be generalized to
a more general class of RSs ${\cal S}_\mu$  where the normal ordering is performed with respect to a generic scale $\mu\neq m$
(in this notation the scheme ${\cal S}$ is identified with ${\cal S}_{\mu=m}$).
In this section we determine the critical coupling and the critical exponent $\nu$ for different RSs ${\cal S}_\mu$ in order to study
the RS-dependence and the effectiveness of the Borel resummation, and also to explicitly determine how much the critical coupling is sensitive
to the choice of  RS.  Of course $\nu$, being a direct physical observable, should be RS-independent, so its evaluation in different RSs
provides also a consistency check of the results. We define a family of schemes parametrised by
\be
\kappa \equiv \log \frac{\mu^2}{m^2}\,,
\ee
and study the dependence on $\kappa$ of the various observables.
For $\kappa\neq 0$ the one-loop tadpole diagram no longer vanishes:
\begin{equation}
	\raisebox{-.33\height}{\includegraphics[width=1.7cm]{img/2pt-1t.pdf}}
	+ \raisebox{-.42\height}{\includegraphics[width=1.3cm]{img/deltam.png}}
	= \frac{3}{\pi} \kappa g \,,
	\label{eq:MSmGamma}
\end{equation}
and hence all loop diagrams involving tadpoles cannot be neglected. Luckily enough, there is no need to compute such diagrams.
\begin{figure}[t]
	\begin{equation*}
		\raisebox{2pt}{\includegraphics[scale=.65]{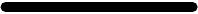}} \longrightarrow
		\raisebox{2pt}{\includegraphics[scale=.65]{img/2pt-prop.pdf}} +
		\includegraphics[scale=.65]{img/2pt-1t.pdf} +
		\includegraphics[scale=.65]{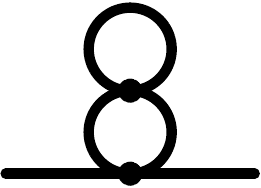} +
		\includegraphics[scale=.65]{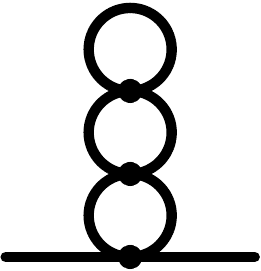} +
		\includegraphics[scale=.65]{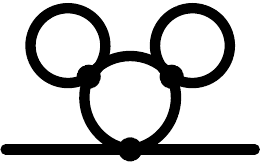} + \dots
	\end{equation*}
	\caption{The repeated application of eq.~\eqref{m2ExpTad} to the 1PI propagator $p^2+m^2$ produces a new series in $\lambda$ which accounts for the bubble diagrams. Each term is regularized as in eq.~\eqref{eq:MSmGamma}.}
	\label{fig:m2ExpTadRepeated}
\end{figure}
If we denote by
\be
O\sim \sum_{k=0}^\infty O_k \Big(\frac{\lambda}{m^2}\Big)^k
\label{eq:ArSer}
\ee
the asymptotic perturbative series for an arbitrary observable $O$ (normalized to make it dimensionless) computed in the RS  ${\cal S}_m$, its perturbative expansion
in the RS ${\cal S}_\mu$ is obtained by iteratively replacing the mass $m^2$ in eq.~(\ref{eq:ArSer}) with
\be
m^2 \rightarrow m^2 + \frac{3\lambda}{\pi} \log \bigg( \frac{\mu^2}{m^2+\frac{3\lambda}\pi \log (\mu^2/(m^2+...))}
\bigg) \,.
\label{m2ExpTad}
\ee
The expansion (\ref{m2ExpTad}) applied to the tree-level term $m^2$ produces all higher-loop bubble diagrams, see fig.~\ref{fig:m2ExpTadRepeated} for their form up to three-loop level.
The replacement (\ref{m2ExpTad}) applies also on possible $m^2$-dependent terms in the coefficients $O_k$ and in the normalization of $O$. E.g. for an $n$-point Schwinger function depending on $x_i$ ($i=1,\ldots, n)$, by dimensional analysis the $c_n$ would generally be functions of $m x_i$. Reexpanding in $\lambda/m^2$ gives the desired perturbative series:
\be
O(\kappa) \sim \sum_{k=0}^\infty O_k(\kappa) \Big(\frac{\lambda}{m^2}\Big)^k\,,
\ee
where clearly $O_k(0)=O_k$. The coefficients $O_k(\kappa)$ are hence uniquely determined from the $O_k$ entering eq.~(\ref{eq:ArSer}).
If $O$ is a direct physical observable, such as a critical exponent, the $\kappa$-dependence should eventually cancel, though in a truncated series a residual dependence would remain.
Note that this procedure applies for all observables, but the vacuum energy and observables related to it, where additional divergences require
further modifications (see e.g. refs.\cite{Serone:2019szm,Rychkov:2015vap} for details).

\subsection{Large Order Behavior}
The large order behavior of the perturbative expansion of $n$-point Schwinger functions $G_n$ in $N$-component $\phi^4$ theories in $2\leq d <  4$ dimensions has been worked out in ref.\cite{Brezin:1992sq} by looking at the semi-classical complex instanton configurations. Following the notation of ref.\cite{Serone:2018gjo},\footnote{Note a typo in eq.~(3.14) of ref.\cite{Serone:2018gjo}. The correct formula should be $b_n = n/2$.} the large order behavior of the coefficients  $G^{(k)}_n$ of the Schwinger functions $G_n$ is given by
\be\label{largeorderNO}
G^{(k)}_n=c_n(-a)^k\Gamma(k+b_n+1)\Big(1+{\cal O}(k^{-1})\Big)\,.
\ee
The knowledge of the coefficients entering eq.~(\ref{largeorderNO}) is valuable when using numerical Borel resummation techniques (in particular the coefficient $a$ is crucial to use the conformal mapping method),
but they will not be needed in the discussion that follows. It is enough for our purposes to know that the coefficients $a$ and $b_n$ are both expected to be RS-independent
while $c_n$ is not \cite{Brezin:1992sq}. It is straightforward to verify this expectation when the $c_n$'s do not depend on $m^2$ using eq.~(\ref{m2ExpTad}).
For large $k$ we find
\be\label{largeordergamma}
G^{(k)}_n(\kappa)= c_n(\kappa) (-a)^k\Gamma(k+b_n+1)\Big(1+{\cal O}(k^{-1})\Big)\,,
\ee
where
\be
c_n(\kappa) = e^{\kappa/a}c_n\,.
\ee
While the choice of RS affects only the overall factor in the large order estimate, the approach to the asymptotic behavior might and indeed does significantly change as the RS is varied.
This is relevant in practice, since we always deal with truncated series.  Let us consider the perturbative expression for the mass gap defined in eq.~(\ref{mphDef}).
Its expression up to order $g^8$ has been found in ref.\cite{Serone:2018gjo} for $\kappa=0$.
Using eq.~(\ref{m2ExpTad}), we can trivially get the mass gap up to order $g^8$ for a generic value of $\kappa$:\footnote{As mentioned before, the replacement should also be taken in the $m^2$ in the denominator of the left hand side of eq.~(\ref{MGapGenericGamma}).}
{\small{
		\begin{align}\label{MGapGenericGamma}
			\frac{M^2}{m^2}= & \ 1+\frac{3}{\pi}\kappa g-\left(\frac{3}{2}+\frac{9}{\pi^2}\kappa\right)g^2+\left(\frac{9}{\pi}+\frac{63\zeta(3)}{2\pi^3}+\left(\frac{27}{\pi^3}+\frac{9}{2\pi}\right)\kappa+\frac{27}{2\pi^3}\kappa^2\right)g^3 \nn \\
			                 & -\left(14.655869(22)+\frac{27}{2\pi^4}\Big(6+ 5\pi^2 + 14\zeta(3)\Big)\kappa+\frac{27}{2\pi^4}(9+\pi^2)\kappa^2+\frac{27}{\pi^4}\kappa^3\right)g^4 \nn                                                               \\
			                 & +\left(65.97308(43)+51.538171(63)\kappa+ \frac{81}{4\pi^5}\Big(36+17\pi^2+42\zeta(3)\Big)\kappa^2+\frac{81}{2\pi^5}(11+\pi^2)\kappa^3+\frac{243}{4\pi^5}\kappa^4\right)g^5\nn                                        \\
			                 & -\left(347.8881(28)+301.2139(16)\kappa+114.49791(12)\kappa^2+\right. \nn                                                                                                                                             \\
			                 & \left.\frac{81}{2\pi^6}\Big(105+37\pi^2+84\zeta(3)\Big)\kappa^3+\frac{243}{4\pi^6}(25+2\pi^2)\kappa^4+\frac{729}{5\pi^6}\kappa^5\right)g^6 \nn                                                                       \\
			                 & +\bigg(2077.703(36)+1948.682(14)\kappa+828.4327(39)\kappa^2+205.20516(19)\kappa^3+ \nn                                                                                                                               \\
			                 & \left.\frac{243}{8\pi^7}\Big(675+197\pi^2+420\zeta(3)\Big)\kappa^4+\frac{729}{20\pi^7}(137+10\pi^2)\kappa^5+\frac{729}{2\pi^7}\kappa^6\right)g^7 \nn                                                                 \\
			                 & -\bigg(13771.04(54)+13765.22(21)\kappa+6373.657(40)\kappa^2+1778.1465(75)\kappa^3+323.93839(27)\kappa^4+\nn                                                                                                          \\
			                 & \left.\frac{2187}{20\pi^8}\Big(812+207\pi^2+420\zeta(3)\Big)\kappa^5+\frac{2187}{20\pi^8}(147+10\pi^2)\kappa^6+\frac{6561}{7\pi^8}\kappa^7\right)g^8+\mathcal{O}(g^9)\,.
		\end{align}
	}}
\begin{table}[t!]
	\centering
	\begin{tabular}{c|c c c c c c}
		\hline
		$\kappa$ & \multicolumn{6}{c}{Loop Order}                                                \\
		         & 3                              & 4       & 5       & 6      & 7      & 8      \\\hline
		-4       & -3.1927                        & -5.0170 & -1.6675 & 4.1456 & 0.6531 & 1.2648 \\
		-3       & -3.0862                        & 5.3535  & 0.6678  & 1.1928 & 1.0617 & 1.0789 \\
		-2       & -0.7251                        & 1.4431  & 1.0521  & 1.0513 & 1.0331 & 1.0309 \\
		-1       & 0.7251                         & 1.0847  & 0.9993  & 0.9825 & 0.9791 & 0.9829 \\\hline
		0        & 1.0040                         & 0.9531  & 0.9113  & 0.9076 & 0.9158 & 0.9284 \\\hline
		1        & 0.9665                         & 0.8468  & 0.8232  & 0.8311 & 0.8489 & 0.8695 \\
		2        & 0.8712                         & 0.7535  & 0.7423  & 0.7585 & 0.7830 & 0.8097 \\
		3        & 0.7767                         & 0.6736  & 0.6711  & 0.6925 & 0.7214 & 0.7521 \\
		4        & 0.6947                         & 0.6064  & 0.6095  & 0.6341 & 0.6653 & 0.6983 \\\hline
	\end{tabular}
	\caption{The ratio of ratios $R_M^{(k)}(\kappa)$ as given by eq.~(\ref{RatioofRatios}) for different values of $\kappa$ and of the loop order $k$.}
	\label{tab:AsymptBehaGamma}
\end{table}
We can see from eq.~(\ref{MGapGenericGamma}) that the coefficient multiplying the $g^n$ term is a polynomial of degree $n-1$ in $\kappa$ for $n>1$. The ${\cal O}(\kappa^{n-1})$ term is determined by the iteration (\ref{m2ExpTad}) and is equal to  $\kappa^{n-1} (-3/\pi)^n \times 1/(1-n)$. Thus more and more low orders terms are dominated
by the ${\cal O}(\kappa^{n-1})$ contribution as $|\kappa|$ gets larger and larger. As a consequence, when $\kappa<0$ many perturbative terms at low order will have the same sign and differ from
the asymptotic estimate (\ref{largeordergamma}).
We compare the ratios of the series of $M^2$ in eq.~(\ref{MGapGenericGamma}) with the ratio of the corresponding asymptotic series for the two-point function $G_2$:
\be\label{RatioofRatios}
R_M^{(k)}(\kappa)=\frac{r^{(k)}_{2,\textnormal{asym}}}{r^{(k)}_{M,\kappa}}\,, \qquad r^{(k)}_{n,\textnormal{asym}}=\frac{G_n^{(k)}}{G_n^{(k-1)}}\,, \quad r^{(k)}_{M,\kappa}=\frac{M^{2(k)}(\kappa)}{M^{2(k-1)}(\kappa)} \,,
\ee
and report $R_M^{(k)}(\kappa)$ for different loop orders $k$ and values of $\kappa$ in table \ref{tab:AsymptBehaGamma}. The behavior described above is evident.
At about $\kappa=-5$ all the terms in eq.~(\ref{MGapGenericGamma}) are positive (apart from the linear term evidently negative). For $\kappa>0$, we see that the alternation of signs is preserved but the deviation from the asymptotic behavior increases with $\kappa$.

\subsection{Mass and critical exponent $\nu$}

We report here the results for the mass gap $M$ and the critical exponent $\nu$ obtained by a numerical Borel resummation, starting from the truncated
expansion (\ref{MGapGenericGamma}). We do not report the details of our numerical implementation. The interested reader can found them in ref.~\cite{Serone:2018gjo}, together with a short introduction to the resummation methods used.

\begin{figure}[t!]
	\includegraphics[width=0.48\textwidth]{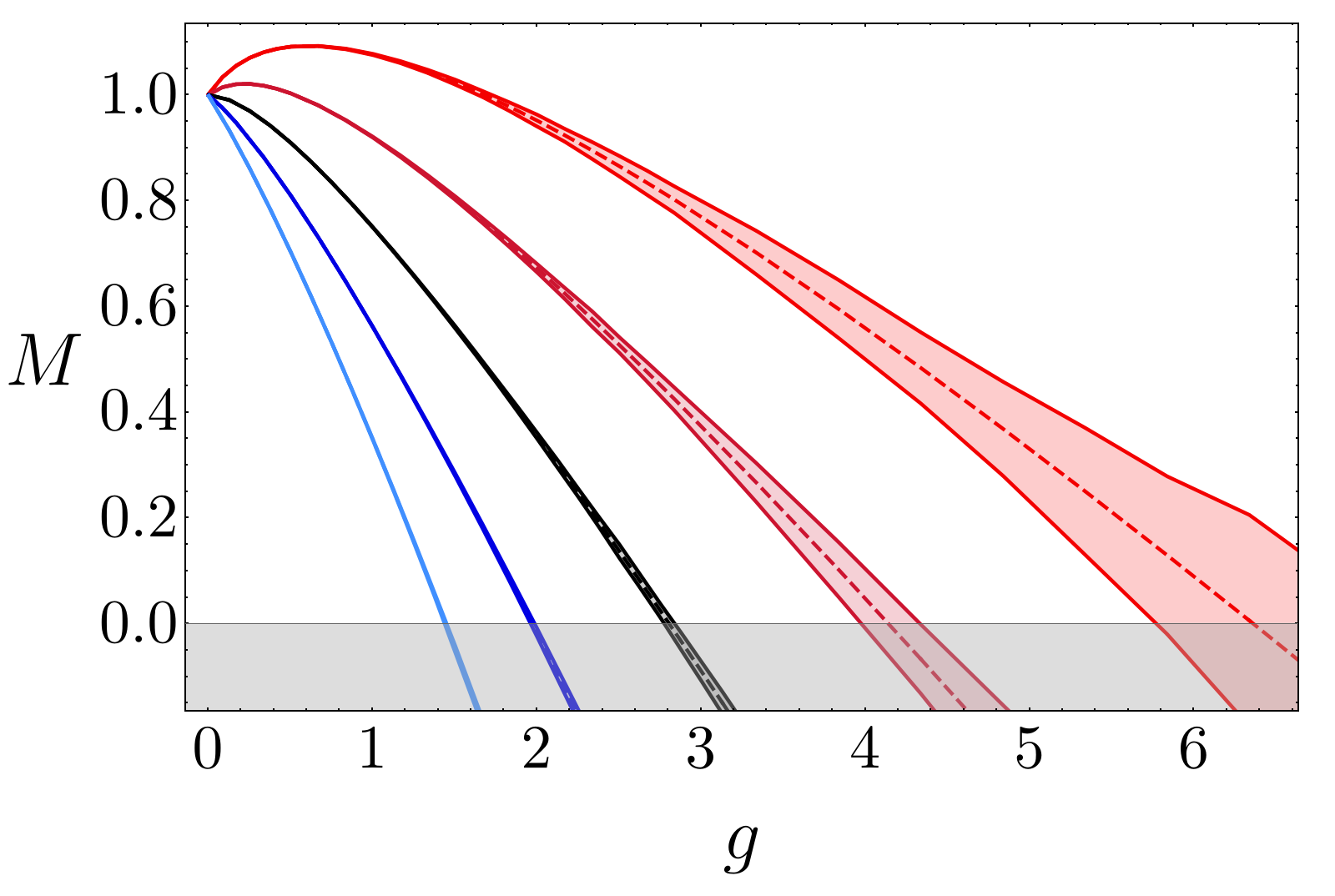} ~
	\includegraphics[width=0.48\textwidth]{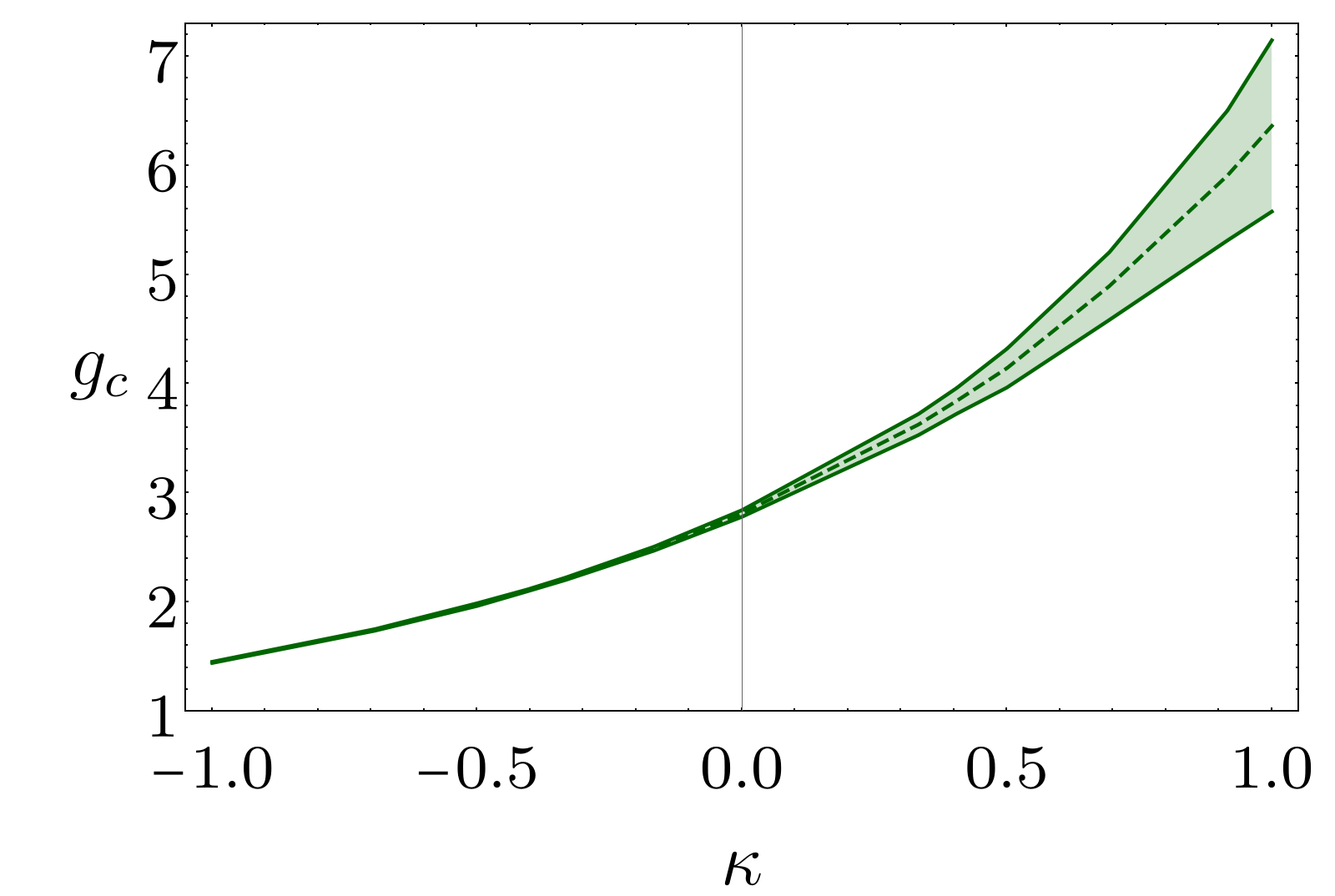}
	\centering
	\caption{(Left panel) The pole mass $M$ as a function of the coupling constant $g$ using conformal mapping for different values of $\kappa$, in order  (from left) $-1,-\frac{1}{2}, 0 \ \textnormal{(in black)} , \frac{1}{2}, 1$. (Right panel) The critical coupling $g_c$, determined as $M(g_c)=0$ using conformal mapping, reported as a function of $\kappa$.}
	\label{fig:Masses_and_gc_of_gamma}
\end{figure}

We show in the left panel of fig.~\ref{fig:Masses_and_gc_of_gamma} the mass gap $M$ as a function of the coupling $g$, for different values of $\kappa$.
All the plots are obtained using the conformal mapping method at order $N=8$. We have verified that similar, but less accurate,
results are obtained using Pad\`e-Borel approximants. For convenience, we also show in the right panel of fig.~\ref{fig:Masses_and_gc_of_gamma}  the value of the critical coupling $g_c$ defined as in eq.~(\ref{gcDef}), as a function of $\kappa$. As can be seen, $g_c$ shows a significant sensitivity on the RS. Moving from $\kappa=0$ to $\kappa=\pm1$ results in a change of $g_c$ of a factor 2.
The value of $g_c$ increases with $\kappa$, in agreement with the naive expectation as dictated by the linear term in $g$ in eq.~(\ref{MGapGenericGamma}). The larger negative values $\kappa$ takes, the smaller $g_c$ becomes, until the critical regime becomes almost accessible in perturbation theory.
Naively one might believe that using a RS with $\kappa \ll -1$ should allow us to get better determinations of the critical regime.
This is however not the case, because for large values of $\kappa$ the tadpole correction (\ref{eq:MSmGamma}) becomes large and perturbation theory unreliable.
The breakdown of perturbation theory is most clear if we take the limit
\be
\kappa\rightarrow - \infty, \quad \quad g\rightarrow 0, \quad \quad {\rm with} \quad g \kappa \equiv y = {\rm fixed}.
\label{eq:limitLargeTad}
\ee
In this limit the mass gap (\ref{MGapGenericGamma}) reduces to
\be
M^2 = m^2 \Big(1 + \frac 3\pi y \Big)\,.
\ee
The critical coupling is predicted to be at $y_c = - \pi/3$ and correspondingly we would analytically get $\nu=1/2$, which corresponds to the mean field theory value, far
from the actual result $\nu=1$.
As we mentioned, the problem arises from the fact that when the log becomes large, the procedure of expanding the $m^2$ terms in the particle propagators
using eq.~(\ref{m2ExpTad}) is no longer justified. In the limit (\ref{eq:limitLargeTad})  we should instead keep in the scalar propagator the one-loop tadpole term,
effectively replacing $m^2$ with $M^2$. In the critical regime where $M\rightarrow 0$ we will then have to face IR divergences that make the perturbative expansion in $g$ (and its resummation) ill-defined.

We now turn to the determination of $\nu$.
For $\kappa\neq 0$, where $M^2$ includes a linear term in $g$, it is useful to resum
\be
L_\kappa(g)\equiv \frac{2g}{g \partial_g \log M^2},
\ee
instead of using eq.~(\ref{eq:Lg}), as in the $\kappa=0$ case \cite{Serone:2018gjo}, and extract $\nu$ as
\be
\nu = \frac{1}{\partial_g L_\kappa}\bigg|_{g=g_c}\,.
\label{nuextrGamma}
\ee
\begin{figure}[t!]
	\includegraphics[width=0.48\textwidth]{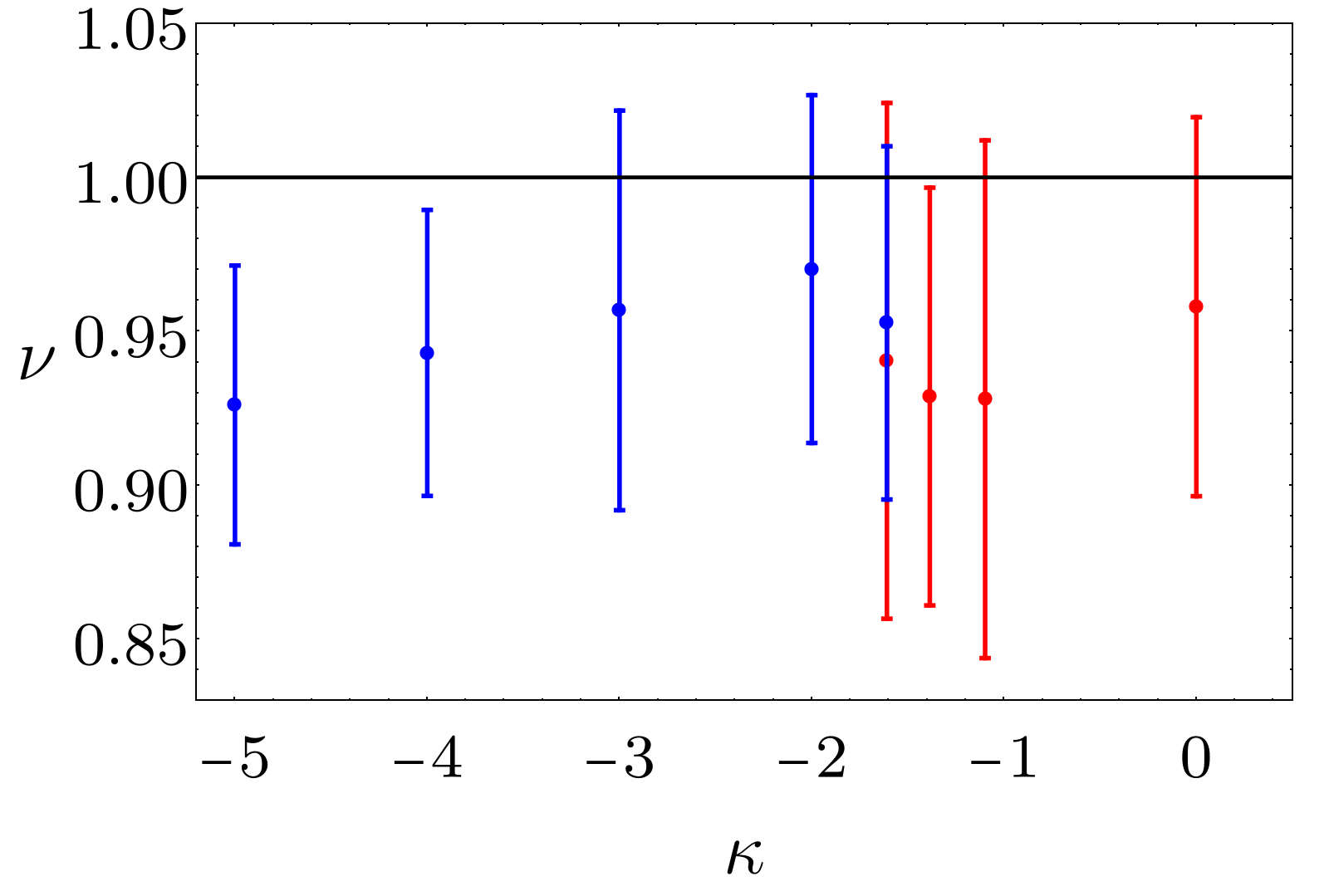} ~
	\includegraphics[width=0.48\textwidth]{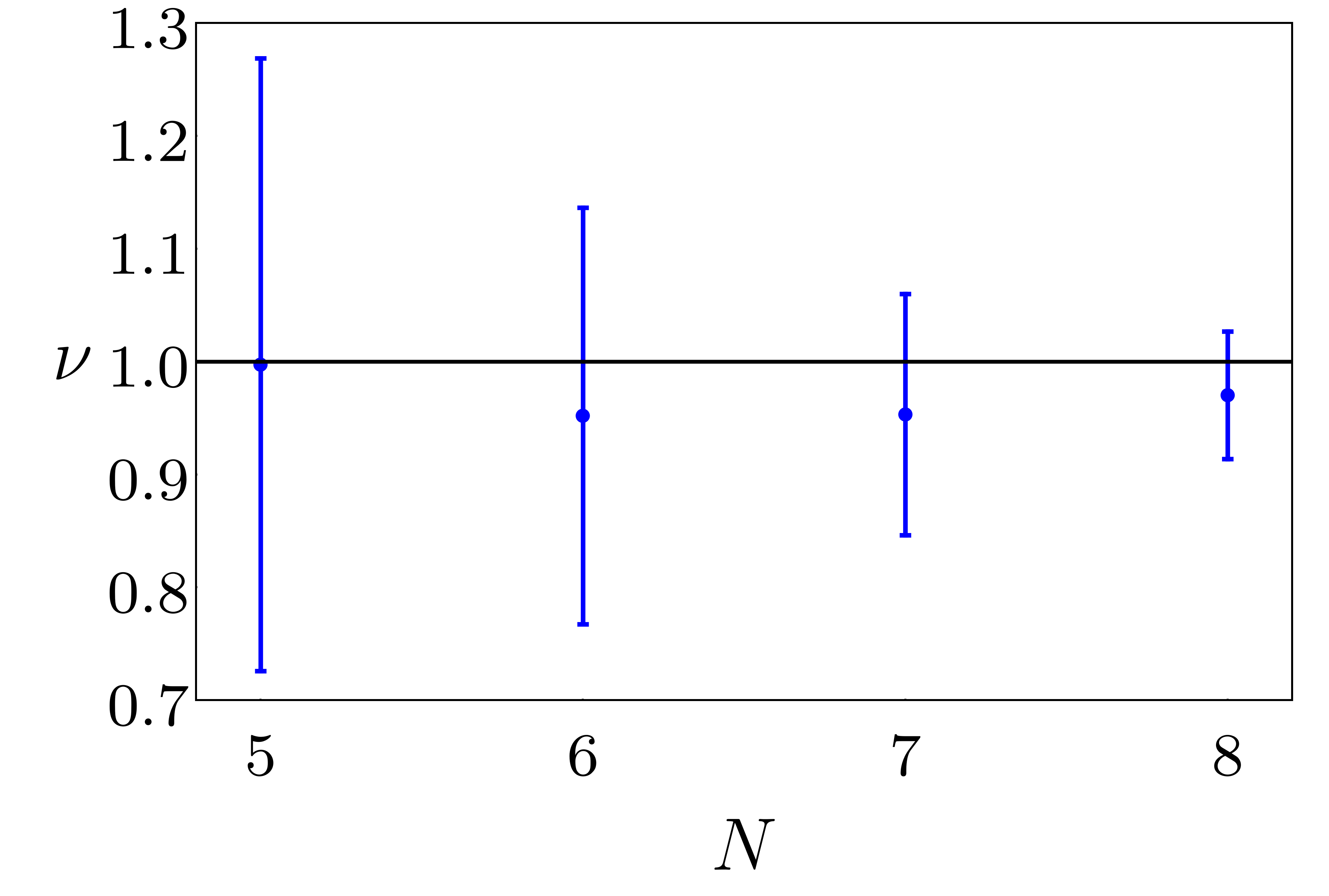}
	\centering
	\caption{(Left panel) The critical exponent $\nu$ for different values of $\kappa$. The blue points are computed with conformal mapping, the red ones with Pad\`e-Borel approximants. (Right panel) The critical exponent $\nu$ is computed with conformal mapping resummation technique for $\kappa=-2$ as function of the number of loops $N$ kept.}
	\label{fig:nu_of_gamma}
\end{figure}
We show in the left panel of fig.~\ref{fig:nu_of_gamma} the values of $\nu$ so determined, as a function of $\kappa$, in the range $\kappa\in [-5,0]$. The resummation using Pad\'e-Borel approximants are not affected by spurious poles only in the range of $\kappa\in [-2,-1]$ and for the value corresponding to the normal ordering RS $\kappa=0$.
The conformal mapping shows an increasingly worse convergence for values of $\kappa>-\frac{3}{2}$, presumably because $L_\kappa(g)$ has a series that differs more and more from the asymptotic one. As can be seen, for the more negative values of $\kappa$ the computed value for $\nu$ starts to depart from its correct value $\nu=1$, drifting towards $\nu=1/2$, as expected from the
previous discussion. We have numerically verified that $\nu\rightarrow 1/2$ as $\kappa\rightarrow -\infty$  if we erroneously continue to resum the perturbative expansion.
The accuracy in the determination of $\nu$ does not significantly change as $\kappa$ is varied in the range where the use of perturbative expansion is justified. For illustration, we show in the right
panel of fig.~\ref{fig:nu_of_gamma} the value of $\nu$ as a function of the coefficient terms kept in the resummation for the value $\kappa=-2$. The improvement as $N$ increases is manifest.

Summarizing, within a specific class of one-parameter family of RSs (normal ordering with $\mu\neq m$) we have quantified how much the large order behavior of the perturbative series for the mass gap and the value of the critical coupling $g_c$ depends on the chosen RS. The value of the critical exponent $\nu$ is instead RS-independent, as it should be.
A spurious RS-dependence arises when a large log hinders perturbation theory and leads to fallacious results.
Within the RSs unaffected by large logs, no significant improvements in the accuracy of the determination of $\nu$ with respect to the standard normal ordering ($\mu=m$) are observed.

\section{Conclusions}

We have pointed out in this paper the importance of the RS in the proof of the Borel summability in $d=2$ and $d=3$ $O(N)$ $\phi^4$ vector models.
In particular, we have shown that the proofs in refs.\cite{Eckmann,Magnen} and ref.\cite{Serone:2018gjo} are both essentially based on the same RS which we denoted by ${\cal S}$.
In this RS, the $\beta$-function of the theory is trivial, and no interesting results can be derived using RG techniques.
On the other hand, most of the results in the literature that make use of the resummation of the fixed dimension expansion (not $\epsilon$-expansion) are based on other RSs, such as
the one we denoted by $\widetilde{\cal S}$ (momentum subtraction) where a non-trivial $\beta$-function occurs. We have shown that in the RS $\widetilde{\cal S}$ the proof  given in ref.\cite{Serone:2018gjo}
no longer holds, and we are not aware of papers in constructive quantum field theory generalising the proofs in refs.\cite{Eckmann,Magnen} to other RSs such as $\widetilde{\cal S}$.

Despite the absence of a non-trivial $\beta$-function, the RS ${\cal S}$ can be used to determine the strong coupling behavior of the theory and its critical regime.
Physical observables such as critical exponents can be extracted from correlation functions or from the behavior of the mass gap as the critical coupling is approached.
The critical exponents so far computed in this way, $\eta$ and $\nu$ in the $d=2$ $\phi^4$ theory \cite{Serone:2018gjo}, are compatible with their exact values, while a long standing mismatch persists in the value of $\eta$ (and of the critical exponent $\omega$) computed in the RS $\widetilde{\cal S}$ by means of eq.~(\ref{eq:etanuRG}). 
These mismatches are believed to be due to some non-analyticities in $\widetilde \beta$ \cite{Pelissetto:1997gk,Calabrese:2000dy}, but we cannot exclude that they signal a problem of Borel summability in the RS $\widetilde{\cal S}$ in the $d=2$ $\phi^4$ theory. From a more practical point of view, in the RS ${\cal S}$ there is no need to compute the 1PI 4-point function $\Gamma^{(4)}$, necessary to determine the $\beta$-function when RG techniques are used.
This is a significant simplification, because at fixed order the number of Feynman diagrams that has to be computed significantly increases with the order of the $n$-point function.

We have then focused on the $d=2$ $\phi^4$ theory and studied how a change in a one-parameter family of RSs concretely affect the effectiveness of the numerical Borel resummation
at a fixed number of orders. While the critical coupling and the deviations of the low order coefficients from the expected large-order behavior significantly depend on the choice of the RS,
the accuracy on a physical observable (critical exponent $\nu$) is essentially constant. As by-product of our analysis, we have improved the determination of $\nu$ obtained in the RS $\widetilde {\cal S}$ (assuming Borel summability) by adding three more orders in its perturbative expansion. The results are reported in table \ref{tab:nucrit} and are in very good agreement with the exact value $\nu=1$.

On physical grounds, the non-Borel summability of a theory is typically due to the presence of non-perturbative effects, such as real action instantons or renormalon singularities.
These effects are RS-independent, so one would be led to the conclusion that the Borel resummability of Schwinger functions should also be a RS-independent property of a theory.
In the context of  $\phi^4$ theories, there is actually substantial numerical evidence of Borel summability in RSs such as $\widetilde {\cal S}$\footnote{Mainly in $d=3$, due to the above mentioned mismatches in the values of $\eta$ and $\omega$ in $d=2$.}  supporting the above claim based on physical intuition.
It would be interesting to clarify this point and prove (or disprove) the RS-independence of the Borel summability of the perturbative expansion.
Perhaps this analysis could give us hints for a possible semi-classical interpretation of renormalon singularities that appear in presence of marginal couplings, where
we no longer have the luxury of defining RSs where the expansion of counterterms in perturbation theory contains a finite number  of terms.

\section*{Acknowledgments}

We thank P. Calabrese, J. Magnen, P. Putrov, and G. Villadoro for discussions. We also thank  A. Pelissetto and S. Rychkov for comments on the manuscript. G.S. acknowledges support from the Paris Île-de-France region in the framework of DIM SIRTEQ.

\appendix

\section{Resumming $\nu$ using the RS \texorpdfstring{$\widetilde {\cal S}$}{Stilde}}
\label{sec:appendix}

In this appendix we show how we can improve on the known perturbative series  for the critical exponent $\nu$ in the RS $\widetilde {\cal S}$  by using the results of ref.\cite{Serone:2018gjo} and the Callan-Symanzik equation (\ref{eq:CSeqsRSs}) for the $d=2$ $N=1$ $\phi^4$ theory.  We start by taking $n=2$ in eq.~(\ref{eq:CSeqsRSs}) and set the momentum $p$ equal to zero.
Given eq.~(\ref{CSMS2}),  the left-hand side of eq.~(\ref{eq:CSeqsRSs}) is completely determined and we can use that equation
to determine the series expansion in $g$ for $\Gamma^{(2,1)}_0(p=0)$. After this is done we can pass to the RS $\widetilde {\cal S}$ by imposing eqs.~(\ref{RenCond}) and (\ref{Gamma21Def}).
In this way, we are able to determine the series expansion for $Z_{\phi^2}$ in the RS $\widetilde {\cal S}$ and in turn the series for $\widetilde \eta_{\phi^2}$ using eq.~(\ref{eq:anomDimphi2}).
After some algebra we can then get the desired perturbative series for $\widetilde \nu^{-1}$:
\begin{equation}
	\begin{split}
		\widetilde\nu^{-1} (v)  =\, & 2 - \frac{2}{3}v + 0.2160804422 \, v^2 -0.2315656721 \, v^3 + 0.3116912(45) \, v^4  \\
		&- 0.55522(17) \, v^5 +1.1694(15) \, v^6-2.819(11) \, v^7+7.496(81) \, v^8 + \mathcal{O}(v^9)
		\,,
	\end{split}
	\label{nuExp}
\end{equation}
where $v\equiv 9\widetilde g/\pi$.
The first 4(5) coefficients in $\widetilde\nu^{-1}$ agree with the results obtained in ref.~\cite{Baker:1977hp} (\cite{Sokolov:2005xp}), providing
a consistency check on the determination of our coefficients up to ${\cal O}(v^5)$. The remaining three coefficients are new.
For completeness, we report the perturbative expansion of the $\beta$-function $\widetilde \beta(v)$ computed in ref.\cite{Serone:2018gjo} (denoted simply by $\beta$ there):\footnote{
	The coupling in the RS $\widetilde{\cal S}$ was denoted by $g_R$ in ref.\cite{Serone:2018gjo} and normalized differently. We have $g_R = 24 \widetilde g$. The definition of $v$ is however the same.}
\begin{small}
	\begin{equation}
		\frac{\widetilde \beta(v)}{2}  = -v + v^2 - 0.7161736 v^3 + 0.930768(3) v^4 - 1.5824(2) v^5 + 3.2591(9) v^6 - 7.711(5) v^7 + 20.12(9) v^8 + \mathcal{O}(v^9).
		\label{eq:A2}
	\end{equation}
\end{small}
The resummation of $\widetilde \nu ^{-1}$ at the critical value of the coupling $v_c$ determined as $\widetilde\beta (v_c) =0$ gives\footnote{As already found in, e.g., ref.\cite{LeGuillou:1979ixc},  more accurate results are obtained by resumming $\nu^{-1}$ rather than $\nu$. Perhaps this is related to the fact that $\nu^{-1}$ is directly related to the anomalous dimension of the operator $\phi^2$, see eq.~(\ref{eq:etanuRG}), but we have not investigated further the reason behind this behavior.}
\begin{table}[t]
	\centering
	\begin{tabular}{ccl}
		\toprule
		$v_c$     & $\widetilde \nu (v_c)$ &             \\
		\midrule
		$1.80(4)$ & $1.023(33)$            & six loops   \\
		$1.82(4)$ & $1.004(47)$            & seven loops \\
		\bottomrule
	\end{tabular}
	\caption{The critical exponent $\widetilde \nu$ obtained by resummation of the perturbative series in eq.~\eqref{nuExp} at six and seven loops.}
	\label{tab:nucrit}
\end{table}
the values of $\widetilde \nu (v_c)$ reported in tab.~\ref{tab:nucrit} which are
in very good agreement with the exact value $\widetilde \nu (v_c)=1$ . The larger error at seven loops is due to the uncertainty in the determination of the higher order coefficients. We didn't use the $\mathcal{O}(v^8)$ order in the determination of $\widetilde \nu$ because the $\beta$-function is known only up to seven loops ($v^8$ order).


\begin{thebibliography}{10}

	\small

	\bibitem{Pelissetto:2000ek}
	A.~Pelissetto and E.~Vicari,
	``Critical phenomena and renormalization group theory,''
	Phys.\ Rept.\  {\bf 368} (2002) 549
	doi:10.1016/S0370-1573(02)00219-3
	[cond-mat/0012164].


	\bibitem{Wilson:1971dc}
	K.~G.~Wilson and M.~E.~Fisher,
	``Critical exponents in 3.99 dimensions,''
	Phys.\ Rev.\ Lett.\  {\bf 28} (1972) 240.
	doi:10.1103/PhysRevLett.28.240

	\bibitem{Parisi:1993sp}
	G.~Parisi,
	``Field theoretic approach to second order phase transitions in two-dimensional and three-dimensional systems,''
	J.\ Stat.\ Phys.\  {\bf 23} (1980) 49.
	doi:10.1007/BF01014429

	\bibitem{Eckmann}
	J.P.  Eckmann, J. Magnen and R. S\'en\'eor,
	``Decay properties and borel summability for the Schwinger functions in $P(\Phi)_2$ theories",
	Commun.Math. Phys.  39, 251 (1975).
	doi:10.1007/BF01705374

	\bibitem{Magnen}
	J.~Magnen and R.~S\'en\'eor,
	``Phase Space Cell Expansion and Borel Summability for the Euclidean $\phi^4$ in Three-Dimensions Theory,''
	Commun.\ Math.\ Phys.\  {\bf 56} (1977) 237.
	doi:10.1007/BF01614211

	\bibitem{Eckmann:1979pr}
	J.~P.~Eckmann and H.~Epstein,
	``Borel Summability Of The Mass And The S Matrix In Phi**4 Models,''
	Commun.\ Math.\ Phys.\  {\bf 68} (1979) 245.
	doi:10.1007/BF01221126

	\bibitem{Serone:2018gjo}
	M.~Serone, G.~Spada and G.~Villadoro,
	``$\lambda \phi^4$ Theory I: The Symmetric Phase Beyond NNNNNNNNLO,''
	JHEP {\bf 1808} (2018) 148
	doi:10.1007/JHEP08(2018)148
	[arXiv:1805.05882 [hep-th]].

	\bibitem{Serone:2019szm}
	M.~Serone, G.~Spada and G.~Villadoro,
	``$\lambda \phi^4$ Theory II: The Broken Phase Beyond NNNN(NNNN)LO,''
	arXiv:1901.05023 [hep-th].

	\bibitem{Brezin:1976vw}
	E.~Brezin, J.~C.~Le Guillou and J.~Zinn-Justin,
	``Perturbation Theory at Large Order. 1. The phi**2N Interaction,''
	Phys.\ Rev.\ D {\bf 15} (1977) 1544.
	doi:10.1103/PhysRevD.15.1544

	\bibitem{Kompaniets:2017yct}
	M.~V.~Kompaniets and E.~Panzer,
	``Minimally subtracted six loop renormalization of $O(n)$-symmetric $\phi^4$ theory and critical exponents,''
	Phys.\ Rev.\ D {\bf 96} (2017) no.3,  036016
	doi:10.1103/PhysRevD.96.036016
	[arXiv:1705.06483 [hep-th]].

	\bibitem{GellMann:1954fq}
	M.~Gell-Mann and F.~E.~Low,
	``Quantum electrodynamics at small distances,''
	Phys.\ Rev.\  {\bf 95} (1954) 1300.
	doi:10.1103/PhysRev.95.1300

	\bibitem{Schloms:1989rq}
	R.~Schloms and V.~Dohm,
	``Minimal Renormalization Without Epsilon Expansion: Critical Behavior in Three-dimensions,''
	Nucl.\ Phys.\ B {\bf 328} (1989) 639.
	doi:10.1016/0550-3213(89)90223-X

	\bibitem{Schloms:1990zz}
	R.~Schloms and V.~Dohm,
	``Minimal renormalization without epsilon expansion: Critical behavior above and below Tc,''
	Phys.\ Rev.\ B {\bf 42} (1990) 6142.
	doi:10.1103/PhysRevB.42.6142

	\bibitem{Callan:1970yg}
	C.~G.~Callan, Jr.,
	``Broken scale invariance in scalar field theory,''
	Phys.\ Rev.\ D {\bf 2} (1970) 1541.
	doi:10.1103/PhysRevD.2.1541

	\bibitem{Symanzik:1970rt}
	K.~Symanzik,
	``Small distance behavior in field theory and power counting,''
	Commun.\ Math.\ Phys.\  {\bf 18} (1970) 227.
	doi:10.1007/BF01649434

	\bibitem{Meneses:2018xpu}
	S.~Meneses, J.~Penedones, S.~Rychkov, J.~M.~V.~P.~Lopes and P.~Yvernay,
	``A structural test for the conformal invariance of the critical 3d Ising model,''
	arXiv:1802.02319 [hep-th].

	\bibitem{Baker:1977hp}
	G.~A.~Baker, Jr., B.~G.~Nickel and D.~I.~Meiron,
	``Critical Indices from Perturbation Analysis of the Callan-Symanzik Equation,''
	Phys.\ Rev.\ B {\bf 17} (1978) 1365.
	doi:10.1103/PhysRevB.17.1365

	\bibitem{LeGuillou:1979ixc}
	J.~C.~Le Guillou and J.~Zinn-Justin,
	``Critical Exponents from Field Theory,''
	Phys.\ Rev.\ B {\bf 21} (1980) 3976.
	doi:10.1103/PhysRevB.21.3976

	\bibitem{Orlov:2000wn}
	E.~V.~Orlov and A.~I.~Sokolov,
	``Critical thermodynamics of the two-dimensional systems in five loop renormalization group approximation,''
	Phys. Solid State (2000) 42: 2151. https://doi.org/10.1134/1.1324056
	[hep-th/0003140].

	\bibitem{Sokolov:2005xp}
	A.~I.~Sokolov,
	``Pseudo-epsilon expansion and the two-dimensional Ising model,''
	Fiz.\ Tverd.\ Tela {\bf 47} (2005) 2056
		[Sov.\ Phys.\ Solid State {\bf 47} (2005) 2144]
	doi:10.1134/1.2131160
	[cond-mat/0510088 [cond-mat.stat-mech]].

	\bibitem{Guida:1998bx}
	R.~Guida and J.~Zinn-Justin,
	``Critical exponents of the N vector model,''
	J.\ Phys.\ A {\bf 31} (1998) 8103
	doi:10.1088/0305-4470/31/40/006
	[cond-mat/9803240].


	\bibitem{Guida:2005bc}
	R.~Guida and P.~Ribeca,
	``Towards a fully automated computation of RG-functions for the 3-d O(N) vector model: Parametrizing amplitudes,''
	J.\ Stat.\ Mech.\  {\bf 0602} (2006) P02007
	doi:10.1088/1742-5468/2006/02/P02007
	[cond-mat/0512222 [cond-mat.stat-mech]].

	\bibitem{Pelissetto:1997gk}
	A.~Pelissetto and E.~Vicari,
	``Four point renormalized coupling constant and Callan-Symanzik Beta function in O(N) models,''
	Nucl.\ Phys.\ B {\bf 519} (1998) 626
	doi:10.1016/S0550-3213(98)00164-3
	[cond-mat/9711078].

	\bibitem{Calabrese:2000dy}
	P.~Calabrese, M.~Caselle, A.~Celi, A.~Pelissetto and E.~Vicari,
	``Nonanalyticity of the Callan-Symanzik Beta function of two-dimensional O(N) models,''
	J.\ Phys.\ A {\bf 33} (2000) 8155
	doi:10.1088/0305-4470/33/46/301
	[hep-th/0005254].

	\bibitem{Milsted:2013rxa}
	A.~Milsted, J.~Haegeman and T.~J.~Osborne,
	``Matrix product states and variational methods applied to critical quantum field theory,''
	Phys.\ Rev.\ D {\bf 88} (2013) 085030
	doi:10.1103/PhysRevD.88.085030
	[arXiv:1302.5582 [hep-lat]].

	\bibitem{Rychkov:2014eea}
	S.~Rychkov and L.~G.~Vitale,
	``Hamiltonian truncation study of the $\phi^4$ theory in two dimensions,''
	Phys.\ Rev.\ D {\bf 91} (2015) 085011
	doi:10.1103/PhysRevD.91.085011
	[arXiv:1412.3460 [hep-th]].


	\bibitem{Bosetti:2015lsa}
	P.~Bosetti, B.~De Palma and M.~Guagnelli,
	``Monte Carlo determination of the critical coupling in $\phi^4_2$ theory,''
	Phys.\ Rev.\ D {\bf 92} (2015) no.3,  034509
	doi:10.1103/PhysRevD.92.034509
	[arXiv:1506.08587 [hep-lat]].

	\bibitem{Pelissetto:2015yha}
	A.~Pelissetto and E.~Vicari,
	``Critical mass renormalization in renormalized $\phi^4$ theories in two and three dimensions,''
	Phys.\ Lett.\ B {\bf 751} (2015) 532
	doi:10.1016/j.physletb.2015.11.015
	[arXiv:1508.00989 [hep-th]].

	\bibitem{Elias-Miro:2017xxf}
	J.~Elias-Miro, S.~Rychkov and L.~G.~Vitale,
	``High-Precision Calculations in Strongly Coupled Quantum Field Theory with Next-to-Leading-Order Renormalized Hamiltonian Truncation,''
	JHEP {\bf 1710} (2017) 213
	doi:10.1007/JHEP10(2017)213
	[arXiv:1706.06121 [hep-th]].

	\bibitem{Elias-Miro:2017tup}
	J.~Elias-Miro, S.~Rychkov and L.~G.~Vitale,
	``NLO Renormalization in the Hamiltonian Truncation,''
	Phys.\ Rev.\ D {\bf 96} (2017) no.6,  065024
	doi:10.1103/PhysRevD.96.065024
	[arXiv:1706.09929 [hep-th]].


	\bibitem{Bronzin:2018tqz}
	S.~Bronzin, B.~De Palma and M.~Guagnelli,
	``New Monte Carlo determination of the critical coupling in $\phi^4_2$ theory,''
	Phys.\ Rev.\ D {\bf 99} (2019) no.3,  034508
	doi:10.1103/PhysRevD.99.034508
	[arXiv:1807.03381 [hep-lat]].

	\bibitem{Kadoh:2018tis}
	D.~Kadoh, Y.~Kuramashi, Y.~Nakamura, R.~Sakai, S.~Takeda and Y.~Yoshimura,
	``Tensor network analysis of critical coupling in two dimensional $\phi^{4}$ theory,''
	arXiv:1811.12376 [hep-lat].


	\bibitem{Serone:2017nmd}
	M.~Serone, G.~Spada and G.~Villadoro,
	``The Power of Perturbation Theory,''
	JHEP {\bf 1705} (2017) 056
	doi:10.1007/JHEP05(2017)056
	[arXiv:1702.04148 [hep-th]].

	\bibitem{Erbin:2019zug}
	H.~Erbin, V.~Lahoche and M.~Tamaazousti,
	``Constructive expansion for quartic vector fields theories. I. Low dimensions,''
	arXiv:1904.05933 [hep-th].

	\bibitem{Jaffe:2000ub}
	A.~M.~Jaffe,
	``Constructive quantum field theory,'' in Mathematical physics 2000, ed.
	A.S. Fokas, A. Grigorian, T. Kibble and B. Zegarlinsky, pp. 111-127.


	\bibitem{Rivasseau}
	V.~Rivasseau,
	``From Perturbative to Constructive Renormalization",
	Princeton University Press, May 1991.

	\bibitem{Hardy}
	G.H. Hardy,
	``Divergent Series", Oxford University Press, 1949.

	\bibitem{Sokal:1980ey}
	A.~D.~Sokal,
	``An Improvement Of Watson's Theorem On Borel Summability,''
	J.\ Math.\ Phys.\  {\bf 21} (1980) 261.
	doi:10.1063/1.524408


	\bibitem{Derrick:1964ww}
	G.~H.~Derrick,
	``Comments on nonlinear wave equations as models for elementary particles,''
	J.\ Math.\ Phys.\  {\bf 5} (1964) 1252.
	doi:10.1063/1.1704233

	\bibitem{Brezin:1976wa}
	E.~Brezin, J.-C.~Le Guillou and J.~Zinn-Justin,
	``Perturbation Theory at Large Order. 2. Role of the Vacuum Instability,''
	Phys.\ Rev.\ D {\bf 15} (1977) 1558.
	doi:10.1103/PhysRevD.15.1558

	\bibitem{Iwaki}
	K.~Iwaki, T.~Nakanishi, ``Exact WKB analysis and cluster algebras", J. Phys. A 47 (2014)
	doi:10.1088/1751-8113/47/47/474009 [arXiv:1401.7094[math.CA]]

	\bibitem{Costin}
	O.~Costin, ``Asymptotics and Borel Summability", Monographs and surveys in pure and applied
	mathematics, vol. 141, Chapmann and Hall/CRC, 2008.

	\bibitem{Auberson:1981xx}
	G.~Auberson and G.~Mennessier,
	``Some Properties of Borel Summable Functions,''
	J.\ Math.\ Phys.\  {\bf 22} (1981) 2472.
	doi:10.1063/1.524806

	\bibitem{Auberson:1985}
	G.~Auberson and G.~Mennessier,
	``The Reciprocal of a Borel Summable Function is Borel Summable,''
	Commun.\ Math.\ Phys.\  {\bf 100} (1985) 439.
	doi:10.1007/BF01206138


	\bibitem{Rychkov:2015vap}
	S.~Rychkov and L.~G.~Vitale,
	``Hamiltonian truncation study of the $\phi^4$ theory in two dimensions. II. The $\mathbb Z_2$ -broken phase and the Chang duality,''
	Phys.\ Rev.\ D {\bf 93} (2016) no.6,  065014
	doi:10.1103/PhysRevD.93.065014
	[arXiv:1512.00493 [hep-th]].

	\bibitem{Brezin:1992sq}
	E.~Brezin and G.~Parisi,
	``Critical exponents and large order behavior of perturbation theory,''
	J. Stat. Phys. 19 (1978) 269-292.


\end{thebibliography}
\end{document}